%% file: intrinsiccharm.tex
\documentclass[a4paper,11pt]{article}
\pdfoutput=1 

\usepackage{jheppub} 
\arxivnumber{1510.02491}

\usepackage[T1]{fontenc} 

\usepackage[latin1]{inputenc}
\usepackage[english]{babel}

\usepackage{amssymb,amsthm,cancel,hyperref,graphicx,xcolor}
\usepackage{picinpar,graphicx,xypic}
\usepackage{booktabs}
\usepackage{mathrsfs}
\usepackage{amsfonts}
\usepackage{latexsym}
\usepackage{booktabs,graphicx,hyperref,epsfig,multirow}
\usepackage{bm}

\def\smallfrac#1#2{\hbox{${{#1}\over {#2}}$}}
\newcommand{\be}{\begin{equation}}
\newcommand{\ee}{\end{equation}}
\newcommand{\beq}{\begin{equation}}
\newcommand{\eeq}{\end{equation}}
\def\bea#1\eea{\begin{align}#1\end{align}}
\newcommand{\MSb}{\overline{\text{MS}}}
\newcommand{\muc}{\mu_c}
\newcommand{\mub}{\mu_b}
\newcommand{\mut}{\mu_t}
\newcommand{\mur}{\mu_{\scriptscriptstyle\rm R}}

\newcommand{\bigGamma}{{\bar\Gamma}}
\newcommand{\as}{\alpha_s}
\newcommand{\Ord}{\mathcal{O}}

\newcommand{\nthree} {{(3)}}
\newcommand{\nthreez}{{(3,0)}}
\newcommand{\nfour}  {{(4)}}
\newcommand{\nfourz} {{(4,0)}}
\newcommand{\nfive}  {{(5)}}
\newcommand{\nfivez} {{(5,0)}}
\newcommand{\nsix}   {{(6)}}
\newcommand{\nsixz}  {{(6,0)}}
\newcommand{\nsf}   {{(6)}}
\newcommand{\nff}   {{(5)}}
\newcommand{\nft}   {{(4)}}

\def\({\left(}
\def\){\right)}
\def\[{\left[}
\def\]{\right]}

\allowdisplaybreaks
\numberwithin{equation}{section}

\preprint{
\begin{flushright}
Edinburgh 2015/06\\
CERN-PH-TH/2015-118\\
OUTP-15-25P
\end{flushright}
}

\title{Charm in Deep-Inelastic Scattering}

\author[a,b]{Richard~D.~Ball,}
\author[c]{Marco Bonvini,}
\author[c]{Luca Rottoli}

\affiliation[a]{The Higgs Centre for Theoretical Physics, University of Edinburgh,\\
  JCMB, KB, Mayfield Rd, Edinburgh EH9 3JZ, Scotland}
\affiliation[b]{PH Department, TH Unit, CERN, CH-1211 Geneva 23, Switzerland}
\affiliation[c]{Rudolf Peierls Centre for Theoretical Physics, 1 Keble Road,\\ University of Oxford, OX1 3NP Oxford, UK}

\abstract 
{We show how to extend systematically the FONLL scheme for inclusion
  of heavy quark mass effects in DIS to account for
  the possible effects of an intrinsic charm component in the nucleon.
  We show that when there is no intrinsic charm, FONLL is equivalent to
  S-ACOT to any order in perturbation theory, while when an intrinsic
  charm component is included FONLL is identical to ACOT, again to all
  orders in perturbation theory. We discuss in detail the inclusion of
  top and bottom quarks to construct a variable flavour number scheme, and give explicit
  expressions for the construction of the structure functions
  $F^c_2$, $F^c_L$ and $F^c_3$ to NNLO.}

\begin{document}
\maketitle
\flushbottom

\clearpage

\input{sections/sec-introduction}

\input{sections/sec-schemes}

\input{sections/sec-fonll}

\input{sections/sec-beauty}

\input{sections/sec-summary}

\appendix

\input{sections/sec-appendixmat}
\input{sections/sec-results}

\phantomsection
\addcontentsline{toc}{section}{References}
\bibliographystyle{jhep}
\bibliography{intrinsiccharm}

\end{document}

%% file: sections/sec-introduction.tex
\section{Introduction}
\label{sec:introduction}

An accurate treatment of heavy quark mass effects is an essential
ingredient of modern PDF
fits~\cite{Ball:2014uwa,Harland-Lang:2014zoa,Nadolsky:2008zw,Alekhin:2013nda,Ball:2015oha}.
Global PDF fits require the computation of physical cross sections
over a range of perturbative scales $Q^2$ in order to incorporate a
wide range of data from fixed target experiments up to LHC. As these
scales pass through (or close to) the thresholds for charm, bottom and
top, precision results require the incorporation of heavy quark mass
effects close to threshold, $Q^2\sim m^2$, and the resummation of
collinear logarithms at scales far above the threshold, $Q^2\gg m^2$,
$m$ being the mass of the heavy quark.  This is achieved through the
use of a so-called variable flavour number scheme (VFNS): calculations involving heavy quarks
in DIS in different schemes with different numbers of
active flavours participating to DGLAP evolution are combined to derive
an expression for the coefficient functions which is valid both close
to threshold, and far above it. A number of such VFNSs have been
proposed for DIS structure functions, including
ACOT~\cite{Aivazis:1993kh,Aivazis:1993pi},
S-ACOT~\cite{Collins:1997sr,Kramer:2000hn},
TR and TR$^\prime$~\cite{Thorne:1997ga,Thorne:2006qt}, and
FONLL~\cite{Buza:1996wv,Cacciari:1998it,Forte:2010ta}.

A common feature of these various VFNSs is that they assume that the
heavy quark PDF is generated entirely perturbatively above threshold.
This assumption is reasonable enough for top and bottom, since both
sit well within the accepted region of validity of perturbative QCD,
and an entirely perturbative treatment is appropriate.  By contrast,
the distribution of gluons, and up, down and strange quarks in the
proton is clearly nonperturbative, and can only be determined
empirically through PDF fits.

The charm quark plays a special role, since the charm threshold sits
at the borderline between perturbative and nonperturbative behaviour.
While at high scales most charm is generated perturbatively through
photo-gluon fusion (so that at HERA for example charm contributes up
to 25\% of the measured structure functions), closer to threshold it is
difficult to rule out a priori a small nonperturbative component.
Ideally one would like to admit the possibility of an initial charm
PDF at threshold, which then evolves perturbatively to higher
scales. The initial charm PDF could then be determined by fitting to
data, just like the gluon and light quark PDFs.  While over the years
a variety of nonperturbative models of this `intrinsic charm' have
been proposed~\cite{Brodsky:1980pb,Vogt:1994zf,Pumplin:2005yf}, and
various attempts have been made at an empirical
determination~\cite{Aubert:1981ix,Harris:1995jx,Dulat:2013hea,Jimenez-Delgado:2014zga},
so far no conclusive evidence has been found.

In this paper we will construct a VFNS which can incorporate intrinsic
heavy quark PDFs, specifically intrinsic charm. We will take as given
the existence to all orders in perturbation theory of the usual
massless $\overline{\rm MS}$ factorization, and the complementary
massive factorization proven in \cite{Collins:1998rz}.  We then
compare the ACOT and FONLL constructions, all the time taking
into account the possibility of an intrinsic component of the charm
PDF. We find in this way that for a specified renormalization and
factorization scheme (namely $\overline{\rm MS}$), the
FONLL~\cite{Buza:1996wv,Cacciari:1998it} and
ACOT~\cite{Aivazis:1993kh,Aivazis:1993pi} constructions give formally identical
results, to all orders in perturbation theory. Moreover, in the limit
of vanishing intrinsic charm, the original FONLL
procedure~\cite{Forte:2010ta} is precisely equivalent to the S-ACOT
prescription~\cite{Kramer:2000hn, Guzzi:2011ew}, again
to all orders in perturbation theory: the only difference between them
is in formally subleading terms implemented through a damping factor
(FONLL) or a phenomenological $\chi$-rescaling (S-ACOT-$\chi$,
\cite{Tung:2001mv}), which parametrize subleading ambiguities in the
implementation of the condition of zero intrinsic charm. The TR
prescription in its original formulation \cite{Thorne:1997ga} was only
specified at NLO, while at NNLO \cite{Thorne:2006qt} it essentially
reduces to S-ACOT \cite{Thorne:2008xf}: as far as we are aware there
is no formal extension to all orders, so this prescription will not be
considered further here.

The basic formalism of the schemes used for fixed order and resummed
results and their matching is developed in Sect.~\ref{sec:fact}.
Then in Sect.~\ref{sec:fonll} we present the FONLL formalism,
we show the formal equivalence of FONLL and ACOT, the simplifications
evident in the limit of no intrinsic charm, and in particular show
that when all charm is generated perturbatively, FONLL is equivalent
to S-ACOT. The inclusion of top and bottom quarks is discussed in Sect.~\ref{sec:bt}.
Conclusions are drawn in Sect.~\ref{sec:conclusion}.
In the Appendices we derive some technical results on matrix inversion, and write
down explicit results for the structure functions $F_2^c$, $F_L^c$ and
$F_3^c$ to NNLO.

%% file: sections/sec-schemes.tex
\section{Heavy Quarks and Factorization}
\label{sec:fact}

The definition of light and heavy quarks is somewhat arbitrary:
being `light' or `heavy' is a relative concept.
In the context of initial state factorization, a convenient definition of `light' quark
is a quark whose mass $m\lesssim\Lambda_{\rm QCD}$,
such that a perturbative treatment is not applicable.
According to this definition, the up, down and strange quarks are light.
Light quarks can be taken to be massless, because the factorization theorem
is accurate up to $\Ord(\Lambda_{\rm QCD}^2/Q^2)$ corrections,
and light quark mass corrections are higher twist effects $\Ord(m^2/Q^2)$ .
Consistently, it is natural to define as `heavy' a quark whose mass $m\gg\Lambda_{\rm QCD}$,
such that $\as(m^2)$ is in the perturbative regime.
With this definition, the bottom and top quarks are heavy, and their description
can be carried out using perturbation theory.

The charm mass $m_c$ sits somewhere around the boundary of the region of validity of perturbative QCD:
if we denote the initial scale of perturbative parton evolution by $Q_0$, such that for $Q>Q_0$ evolution is perturbative,
while for $Q<Q_0$ nonperturbative behaviour sets in, then $Q_0\sim m_c$.
For this reason, the charm is special, since it is not heavy enough to fully trust perturbation theory,
but not light enough that its mass can be ignored.
Therefore, we cannot safely assume that at $Q_0$ the charm distributions $c(x,Q_0)$ and $\bar c(x,Q_0)$ are strictly zero,
even if $Q_0$ is below the threshold for perturbative charm production,
since nonzero distributions (commonly called `intrinsic charm') may be generated by nonperturbative effects.
To take this into account we need to treat charm in the same way that we treat the light partons $q = u,d,s$,
$\bar{q} = \bar{u},\bar{d},\bar{s}$ and $g$, with an initial (fitted) PDF at $Q_0$,
evolved up to scales $Q>Q_0$ using perturbation theory.
However, unlike the other light partons, charm mass effects cannot be neglected
for scales $Q$ which are not much larger than $m_c$, as typically encountered in DIS experiments.

The computation of coefficient functions can be performed in different factorization and renormalization schemes,
all leading to results for physical cross sections which must be equivalent to all orders in perturbation theory, and must therefore differ at finite order only by higher order corrections.
For renormalization, the quark mass does not play an important role, since a UV divergent massless-quark loop
would be still divergent even if the quark were massive.
Renormalization can be performed in $\MSb$ for all quark families; however,
all flavours would then participate to $\as$ evolution at any scale, resulting in unphysical heavy quark effects at scales much smaller than the heavy quark mass.
It is therefore more appropriate to use $\MSb$ only for quarks with masses lighter than the renormalization scale $\mur$,
and use the CWZ scheme \cite{Collins:1978wz} for quarks heavier than $\mur$, since this scheme, being based on a subtraction at zero external momenta,
ensures decoupling of the heavy quark with mass $m$ for scales $\mur\ll m$.
In this way, resummation of large logarithms of $\mur/m$ with $\mur\gg m$ is achieved,
while the analogous logarithms when $m\gg\mur$ are power suppressed as $\mur^2/m^2$.
Since the choice of which renormalization scheme to use with each quark depends on the relative size
of the quark mass and the renormalization scale, which varies dynamically,
a variable flavour number (renormalization) scheme is generated.

On the other hand, the quark mass acts as an IR regulator.
This means that radiative corrections involving massive quarks are finite.
Thus while for massless quarks factorization is mandatory, for massive quarks
one may choose whether to factorize massive collinear logarithms or not.
In principle, there is nothing wrong with using standard massless factorization for the light quarks,
while keeping massive collinear logarithms in the coefficient functions: this is the so called 3 flavour scheme (3FS),
discussed in Sect.~\ref{sec:massive}.
However, the collinear logarithms (appearing as single logarithms, so there are $k$ logarithms of $Q^2/m^2$ at order $\as^k(Q^2)$)
can become large at high scales, spoiling the perturbative convergence of the 3FS result.
In this case, it is more appropriate to factorize and resum the collinear logarithms
associated in the first place to the charm, then the bottom and at very high scales the top as well, leading to 4, 5 and 6 flavour schemes respectively.
The 4FS is discussed in Sect.~\ref{sec:massless}, with particular emphasis on the charm quark
(since the bottom and the top are treated identically in the 3FS and 4FS).
This will give us the opportunity to discuss the issues
related to a possible intrinsic component of the charm PDF.
The extension of this discussion to the bottom and top quark is straightforward, so 
the details are postponed to Sect.~\ref{sec:bt}.

\subsection{The 3 flavour scheme}
\label{sec:massive}

As already discussed, since the heavy quark mass regulates the IR behaviour,
there is no need to factorize the (finite) collinear logarithms
due to splittings involving the charm, bottom and top quarks.
One can therefore use standard ($\MSb$) massless factorization for the gluon and light quarks only,
and leave explicit collinear logarithms due to massive charm, bottom and top quarks unfactorized in the coefficient function.
Together with the adoption of the decoupling scheme for UV renormalization of charm, bottom and top loops,
this gives the so-called 3 flavour scheme (3FS)~\cite{Collins:1978wz,Collins:1997sr}.
In the context of heavy quark factorization, this is also often called the `massive scheme', since the quark mass dependence of each massive quark, and in particular the charm, is exact. The coefficient functions will then contain unresummed (and at high scales potentially large) mass collinear logarithms.

In this scheme, only the 3 flavours of light quarks (plus the gluon)
evolve with standard DGLAP equations, as a consequence of the massless
$\MSb$ factorization acting only on those flavours.  The contribution
of the charm, bottom and top quarks (both in loops and trees) is
evaluated at fixed order in perturbation theory, without subtraction
and resummation of the related collinear logarithms.  The 3FS is thus
useful in the threshold region of the charm, where in particular the
effects of the charm quark mass are treated explicitly, but breaks
down at higher scales due to large unresummed logarithms of $Q^2/m_c^2$.
Explicitly we then have, for the generic structure
function,\footnote{Throughout this paper we will use consistently a
  superscript $(n)$ to denote a coefficient function in a scheme with
  $n$ active flavours.}
\begin{equation}
  F^{\nthree} (Q^2,m_c^2) =  \sum_{i = g, q, \bar{q}, c ,\bar{c}}
  C_i^{\nthree} \left( \smallfrac{m_c^2}{Q^2}, \alpha_s^{\nthree} (Q^2) \right)\otimes
  f_i^{\nthree} (Q^2) \label{eq:Fthree}
\end{equation}
where $\otimes$ is the usual $x$-space convolution, and we have suppressed all explicit dependence on $x$.
The coefficient functions $C_i^{\nthree}$ include the effects of the charm mass
order by order in perturbation theory,
both in tree diagrams (for example a charm quark emerging from the proton and being struck by a virtual photon)
and in loops (for example a photon-gluon fusion creating a charm-anticharm pair, or a virtual charm loop in a gluon propagator). This dependence includes thresholds: for example the contribution to the coefficient function from photon-gluon fusion includes a factor of $\theta(W^2-4m_c^2)$, to ensure that it vanishes below threshold.
In writing Eq.~\eqref{eq:Fthree} 
we include from the start an explicit contribution from a charm PDF (i.e., the sum runs also over $i=c,\bar c$): if all charm were generated perturbatively we would set this contribution to zero, and the structure function would then depend only on the light PDFs $f_i^{\nthree}$ with $i=g,q,\bar q$.
We write explicitly only the dependence on the charm mass $m_c$, since this is our main focus here,
but we note in passing that the structure function and the coefficient functions can also depend on the bottom and top quark masses as well through virtual loops and, when kinematically allowed, pair production.

The label ${\nthree}$ means that there are only $3$ `active' quarks, by which we mean that they evolve as
\begin{equation}
f^{\nthree}_i(Q^2)= \sum_{j = g, q, \bar{q}}\Gamma^{\nthree}_{ij}\left(Q^2,Q_0^2\right)\otimes f^{\nthree}_j(Q_0^2),
\qquad i=g,q,\bar q,
\label{eq:Ethreesub}
\end{equation}
where $\Gamma^{\nthree}_{ij}\left(Q^2,Q_0^2\right)$ is the solution of the DGLAP equation with three active flavours.
The charm is still present but not active, and in particular the $Q^2$ dependence of the charm contribution to the structure function is all in the coefficient function, so $f^{(3)}_{c ,\bar{c}}$ are independent of $Q^2$ for all $Q^2$.
Since in Eq.~\eqref{eq:Fthree} we have four flavours, even though only three are active, it is convenient to write 
\begin{equation}
f^{\nthree}_i(Q^2)= \sum_{j =g, q, \bar{q},c ,\bar{c}}\bigGamma^{\nthree}_{ij}\left(Q^2,Q_0^2\right)\otimes f^{\nthree}_j(Q_0^2),
\qquad i=g,q,\bar q,c,\bar c,
\label{eq:Ethree}
\end{equation}
where 
\begin{equation}
\bigGamma_{ij}^{\nthree}(Q^2,Q_0^2) = \begin{cases}\Gamma_{ij}^{\nthree}(Q^2,Q_0^2)&\qquad i,j=g, q, \bar{q}\\
\delta_{ij}, &\qquad {i,j=c, \bar{c}}\\ 
0, &\qquad {\rm otherwise.} \end{cases}\label{eq:Gamthree}
\end{equation}

The massive coefficient functions $C_i^{\nthree}(m_c^2/Q^2, \as^{\nthree} (Q^2))$ are computed
to a fixed order in perturbation theory, retaining the full mass dependence of the diagrams,
including in particular the kinematic thresholds arising from the charm quarks in the final state.
They are fully known to $\Ord(\as^2)$~\cite{Laenen:1992zk,Buza:1995ie}
for incoming light partons, but only to $\Ord(\as)$ for incoming heavy partons~\cite{Hoffmann:1983ah,Kretzer:1998ju}.
If there is no initial state (intrinsic) charm, the structure function Eq.~\eqref{eq:Fthree}
does not include contributions from $C_{c,\bar{c}}^{\nthree} ( m_c^2/Q^2, \alpha_s^{\nthree} (Q^2))$. 
In this case, below the threshold for charm pair production the only charm mass effect is through virtual loops;
the adoption of the CWZ renormalization scheme ensures that the charm quark decouples completely below threshold $W^2<m_c^2$
and thus at low scales $Q^2\ll m_c^2$, so that in this limit
\begin{equation}
  F^{\nthree}(Q^2,m_c^2) =
  \sum_{i = g, q,\bar{q}} C_i^{\nthree}\left( \alpha_s^{\nthree}(Q^2) \right)\otimes f_i^{\nthree}(Q^2)
  + \Ord\(\smallfrac{Q^2}{m_c^2}\), \label{eq:Fthreedec}
\end{equation}
where the charm mass dependence has completely disappeared from the coefficient function.
While the heavy limit is clearly not perturbative in the case of charm, it applies equivalently to bottom and top.
Thus for instance the top quark can be ignored if we work at energies far below the top threshold.

\subsection{The 4 flavour scheme}
\label{sec:massless}

In the 3FS the finite mass logarithms arising from the splittings of the charm quark
appear at fixed order in the coefficient functions $C^{\nthree}_i$.
Explicitly, the massive coefficients have a decomposition
\beq
C_i^{\nthree} \( \smallfrac{m_c^2}{Q^2}, \as^{\nthree} (Q^2) \)
=
\sum_{k=0}^\infty \[\as^{\nthree} (Q^2)\]^k \sum_{j=0}^k A_{i,k,j}\(\smallfrac{m_c^2}{Q^2}\) \log^j \smallfrac{m_c^2}{Q^2},
\label{eq:Cthreedecomposition}
\eeq
where the dependence on the logarithms has been made fully explicit,
and the coefficients $A_{i,k,j}(m_c^2/Q^2)$ admit a power expansion on their argument.
At large scales $Q^2\gg m_c^2$, these logarithms become large, eventually spoiling the convergence of the
fixed-order result Eq.~\eqref{eq:Fthree}. In this regime resummation of the collinear logarithms is necessary for reliable predictions.

Therefore, at scales higher than the charm mass, it is advisable (and eventually necessary)
to use a different factorization scheme where these logarithms are factorized into the definition
of the PDFs, and resummed through PDF evolution, just as for the light partons.
This would be mandatory if one considered the charm quark as a massless flavour,
as appropriate in the high energy limit $Q^2\gg m_c^2$: 
in this limit, all collinear divergences including those from charm quarks
have to be subtracted.
Using standard massless $\MSb$ subtractions, the resulting evolution equation reads
\begin{equation}
f^{\nfour}_i(Q^2)=\sum_{j = g, q, \bar{q}, c ,\bar{c}}\Gamma^{\nfour}_{ij}\left(Q^2,Q_0^2\right)\otimes f^{\nfour}_j(Q_0^2)\label{eq:Efour}
\end{equation}
where $\Gamma^{\nfour}_{ij}$ is the DGLAP evolution factor to a given order in perturbation theory for four active (massless) flavours,
resumming all collinear logarithms of $Q^2/Q_0^2$, including those generated by charm splittings.
Collinear logarithms due to bottom and top quarks, however, are not resummed in this scheme,
and will therefore continue to appear at fixed order in the coefficient functions.

We now focus on the computation of the coefficient functions.
In the high energy limit where all four active flavours are considered massless,
we can obtain the structure functions using standard massless $\MSb$ collinear counterterms
also for the charm quark, up to corrections suppressed by powers of $m_c^2/Q^2$.
We thus get
\bea
F^{\nfour} (Q^2,m_c^2) &= F^{\nfour} (Q^2,0) + \Ord\(\smallfrac{m_c^2}{Q^2}\) \nonumber\\
F^{\nfour} (Q^2,0) &= \sum_{i =g, q, \bar{q}, c,
  \bar{c}} C_i^{\nfour} \left(0,\alpha_s^{\nfour} (Q^2) \right)\otimes
f_i^{\nfour} (Q^2), \label{eq:Fzero}
\eea
where $C_i^{\nfour}(0, \alpha_s^{\nfour} (Q^2))$ are the usual massless scheme coefficient functions, analogous to the $C_i^{\nthree}$ of Eq.~\eqref{eq:Fthree} but with an additional massless quark,
evaluated to the given order in perturbation theory in the four flavour running coupling $\as^{\nfour}(Q^2)$.
Here, the first argument has been set to zero to remind us that the charm mass has been neglected
(while the bottom and top masses are finite).
These massless coefficient functions have been computed to $\Ord(\as^3)$~\cite{Vermaseren:2005qc}.
Note that, while $C_i^{\nthree}(m_c^2/Q^2, \as^{\nthree})$ is logarithmically divergent when $m_c^2\to 0$, $C_i^{\nfour}(0, \alpha_s^{\nfour})$
is finite, due to the subtraction of the collinear divergences.

While Eq.~\eqref{eq:Fzero} is acceptable at high scales where the corrections
of $\Ord(m_c^2/Q^2)$ are negligible, it is not legitimate for lower scales closer to the charm mass. In order to make it valid at all scales, the neglected power corrections must be reinstated, at least at fixed order (and this is sufficient, because at high scales they vanish faster than
the growth of the logarithms).
This is the approach adopted in the FONLL prescription, and also
in the recently proposed derivation of Ref.~\cite{Bonvini:2015pxa}.
Once this is done, and the missing power corrections are expressed in terms of the same 4 flavour PDFs evolving
as in Eq.~\eqref{eq:Efour}, we must have a factorized result of the form
\begin{equation}
 F^{\nfour} (Q^2,m_c^2) = \sum_{i = g, q,\bar{q}, c, \bar{c}} 
  C_i^{\nfour} \left(\smallfrac{m_c^2}{Q^2},\alpha_s^{\nfour} (Q^2) \right)\otimes
  f_i^{\nfour} (Q^2),  \label{eq:Ffour}
\end{equation}
where $C_i^{\nfour}(m_c^2/Q^2, \as^{\nfour})$ are coefficient functions which include the effects of the charm mass.
As we shall see later in Sect.~\ref{sec:fonll}, the exact form of the mass-dependent part of these coefficient functions
is not uniquely fixed in the case of perturbatively generated charm: this has led to the construction
of several different (though equally valid) formulations in the literature,
such as ACOT~\cite{Aivazis:1993kh,Aivazis:1993pi},
S-ACOT~\cite{Collins:1997sr,Kramer:2000hn},
TR and TR$^\prime$~\cite{Thorne:1997ga,Thorne:2006qt}, 
FONLL~\cite{Buza:1996wv,Cacciari:1998it,Forte:2010ta}, and the recent formulation of Ref.~\cite{Bonvini:2015pxa}.

A particular form of the coefficient functions, which does not depend on any assumptions about intrinsic charm, is the one obtained in the ACOT scheme \cite{Aivazis:1993pi}, which uses a special factorization scheme the existence of
which has been proved to all orders in perturbation theory by Collins~\cite{Collins:1997sr}. These coefficient functions are obtained by using massless collinear counterterms for the light partons, and massive 
collinear counterterms (using the quark mass as an infrared regulator) for the charm quark, and then applying the usual subtraction procedure
while keeping charm mass dependence everywhere.
The resulting anomalous dimensions correspond to the DGLAP anomalous dimensions,
and lead therefore to the same evolution Eq.~\eqref{eq:Efour}.
Hence, the Collins (ACOT) result can be interpreted as a massive extension of the massless $\MSb$ factorization scheme.
We will regard at Eq.~\eqref{eq:Ffour} as the result obtained in this scheme.

Note that, since in both Eq.~(\ref{eq:Fzero}) and Eq.~(\ref{eq:Ffour}) all collinear singularities
are factorized into the PDFs Eq.~(\ref{eq:Efour}), then if $Q^2 \gg m_c^2$,
\beq\label{eq:C4limit}
C_i^{\nfour} \left(\smallfrac{m_c^2}{Q^2},\alpha_s^{\nfour} (Q^2) \right)
= C_i^{\nfour}\(0,\alpha_s^{\nfour} (Q^2)\) + \Ord\(\smallfrac{m_c^2}{Q^2}\).
\eeq
It can be observed that, since the coefficient functions
$C^{\nfour}_i (m_c^2/Q^2, \as^{\nfour}(Q^2) )$
contain the correct mass dependence and do not contain
mass logarithms, the result Eq.~\eqref{eq:Ffour} performs the collinear resummation
of charm massive logarithms as the massless result Eq.~\eqref{eq:Fzero},
but it additionally includes the exact charm mass dependence as the massive result Eq.~\eqref{eq:Fthree}.
In this respect, this result already provides a satisfactory treatment of heavy quarks in the initial state.
Practically, however, it has never been used beyond NLO, due to complications in the computation
of the massive coefficient functions $C^{\nfour}_i (m_c^2/Q^2, \as^{\nfour}(Q^2) )$.

\subsection{Matching}
\label{sec:matching}

The two results Eqs.~\eqref{eq:Fthree} and \eqref{eq:Ffour} are alternative expressions for the same structure function,
written in terms of different ingredients, specifically $\as$ and the PDFs.
It is the purpose of this section to relate these ingredients in the two schemes.

This proceeds in two stages: first we must match the two renormalization and factorization schemes
at some matching scale $\muc^2\sim m_c^2$, and then we evolve to $Q^2$. For the running coupling
this gives the relation
\begin{equation}
\alpha_s^{\nthree} (\muc^2)  =  \alpha_s^{\nfour} (\muc^2) +
\sum^{\infty}_{p = 2} a_p\left( \alpha_s^{\nfour} (\muc^2) \right)^p  
  ,  \label{eq:matchalpha}
\end{equation}
with coefficients $a_p$ that are readily computed order by order, and
are known up to four loops~\cite{Chetyrkin:1997sg}.
The relation between $\alpha_s^{\nthree}$ and $\alpha_s^{\nfour}$ at the generic scale $Q^2$
can be obtained using renormalization group evolution from $\muc^2$ to $Q^2$.
Given these coefficients, we can choose to expand any perturbative quantity either
in powers of $\alpha_s^{\nthree}$ or in terms of $\alpha_s^{\nfour}$,
by using the relation Eq.~\eqref{eq:matchalpha} or its inverse. Clearly
when comparing coefficients in perturbative expansions, it is
necessary to expand all quantities consistently. In what follows we
will leave all $\alpha_s$ dependence implicit, reinstating it only
when we perform explicit perturbative expansions in Appendix B.

For the factorization the matching condition is likewise
\begin{align}
  f_i^{\nfour} (\muc^2) 
  &= \sum_{j = g, q, \bar{q}, c, \bar{c}} K_{ij}\left( \smallfrac{m_c^2}{\muc^2}\right)\otimes f_j^{\nthree} (\muc^2)
     ,\label{eq:matchpdf}\\
  K_{ij} \left( \smallfrac{m_c^2}{\muc^2}\right)
  &= \delta_{ij}+\sum^{\infty}_{p = 1}\left( \alpha_s^{\nfour} (\muc^2) \right)^p K^{p}_{ij}\left( \smallfrac{m_c^2}{\muc^2}\right)
     , \label{eq:matchpdfexp4}
\end{align}
where the coefficients $K^p_{ij}(m_c^2/\muc^2)$ are
determined perturbatively, by requiring that the 3FS result Eq.~\eqref{eq:Fthree} and
the 4FS result Eq.~\eqref{eq:Ffour} are equal order by order in (the same) $\as$.
The computation can be simplified by taking the massless limit:
inserting Eq.~\eqref{eq:matchpdf} in Eq.~\eqref{eq:Fzero},
we recover Eq.~\eqref{eq:Fthree} up to power suppressed contributions provided
\begin{equation}
\sum_{i = g, q, \bar{q}, c, \bar{c}} C_i^{\nfour}(0) \otimes K_{ij} \left( \smallfrac{m_c^2}{Q^2}\right)
= C_j^{\nthreez} \left(\smallfrac{m_c^2}{Q^2}\right)
\label{eq:masslimdef}
\end{equation}
where in the right hand side $C_j^{\nthreez}$ is just $C_j^{\nthree}$, but with all power suppressed contributions be set to zero,
keeping only mass independent terms and the mass logarithms.
Using the explicit form Eq.~\eqref{eq:Cthreedecomposition}, we can write exactly
\beq
\sum_{i = g, q, \bar{q}, c, \bar{c}} C_i^{\nfour}(0) \otimes K_{ij} \left( \smallfrac{m_c^2}{Q^2}\right)
= 
\sum_{k=0}^\infty \(\as^{\nfour} (Q^2)\)^k \sum_{l=0}^k A_{j,k,l}(0) \log^l \smallfrac{m_c^2}{Q^2}
\label{eq:matchdef1}
\eeq
where the power suppressed contributions have been removed by computing the coefficients $A_{j,k,l}$
for $m_c=0$.
Thus, the matching coefficents $K_{ij}(m_c^2/\muc^2)$
depend on its argument through the logarithms $\log (m_c^2/\muc^2)$ which are present and unresummed in the 3FS coefficients $C^{\nthree}_i(m_c^2/Q^2)$.
Inverting Eq.~\eqref{eq:masslimdef} we can write
\beq
\lim_{m^2_c\to 0}\sum_{j = g, q, \bar{q}, c, \bar{c}}
  C_j^{\nthree} \left(\smallfrac{m_c^2}{Q^2}\right)\otimes
  K_{ji}^{-1} \left( \smallfrac{m_c^2}{Q^2}\right)
=C_i^{\nfour}(0) .\label{eq:matchdef} 
\eeq
which shows that $K_{ij}^{-1}(m_c^2/Q^2)$ factor out the potentially large logarithms
from the massive 3FS coefficient functions in order to ensure the correct massless
limit, where all collinear logarithms have been cancelled.

In practice the matching coefficients $K_{ij}(m_c^2/Q^2)$ are computed by
comparing calculations of deep inelastic coefficients functions in the 3FS and 4FS to
a given order in perturbation theory, and using
Eq.~\eqref{eq:masslimdef} or equivalently
Eq.~\eqref{eq:matchdef}.\footnote
{Alternatively, these coefficients can be computed as a matching between two effective theories of QCD,
as described in Ref.~\cite{Bonvini:2015pxa}.} 
The components of $K_{ij}$ with any $i$ (light or heavy) and
$j=g,q,\bar q$ are fully known to $\Ord(\as^2)$~\cite{Buza:1996wv,Bierenbaum:2009zt},
and some of them also to $\Ord(\as^3)$~\cite{Ablinger:2010ty,Ablinger:2014lka,Ablinger:2014vwa,Ablinger:2014nga}.
On the other hand, the components $K_{ic}$ and $K_{i\bar c}$ for any value of $i$ are only known to
$\Ord(\as)$~\cite{Mele:1990cw}.  The off-diagonal components with a
gluon and a heavy quark, namely $K_{cg}$, $K_{\bar cg}$, $K_{gc}$ and
$K_{g\bar c}$, start contributing at $\Ord(\as)$, while all other
off-diagonal components are nonzero only at $\Ord(\as^2)$.  The
diagonal quantities are all of the form $K_{ii}=1+\Ord(\as)$: while
$K_{gg}$ gets a contribution at $\Ord(\as)$ due to a heavy quark loop,
and $K_{cc}=K_{\bar c\bar c}$ are nontrivial at $\Ord(\as)$, the light
quark components get corrections only at $\Ord(\as^2)$.

It is important to realise that the matching condition
Eq.~\eqref{eq:matchpdf} only holds (to any fixed order) at the particular matching scale
$\muc\sim m_c$, otherwise there would be unresummed large logarithms.
The 4FS PDFs $f_j^{\nfour} (Q^2)$ at the generic scale $Q^2>\muc^2$ are then obtained by
evolving up with DGLAP evolution, Eq.~\eqref{eq:Efour},
\begin{equation}
  f_i^{\nfour} (Q^2)  = \sum_{j,k = g,  q, \bar{q}, c, \bar{c}} \Gamma_{ij}(Q^2,\muc^2)\otimes
  K_{jk}\( \smallfrac{m_c^2}{\muc^2}\) \otimes f_k^{\nthree} (\muc^2) ,
  \label{eq:evolved4FSpdf}
\end{equation}
as standard in common VFNS PDF sets.
Note that, to any fixed order, the 4FS PDFs implicitly depend on the charm mass $m_c$ and the charm threshold $\muc$,
though the last dependence is formally higher order.
To transform $f_j^{\nthree} (Q^2)$ to
$f_j^{\nfour} (Q^2)$ we must also evolve to $Q^2$ the 3FS PDFs using
Eq.~\eqref{eq:Ethree}. We thus find that
\begin{equation}
  f_i^{\nfour} (Q^2)  = \sum_{j = g,  q, \bar{q}, c, \bar{c}} T_{ij}(Q^2,\muc^2,m_c^2)\otimes f_j^{\nthree} (Q^2) , \label{eq:transpdf}
\end{equation}
where we introduced the transformation matrix
\begin{equation}
  T_{ij} (Q^2,\muc^2,m_c^2)  = \sum_{k,l = g, q, \bar{q}, c, \bar{c}}\Gamma_{ik}^{\nfour}(Q^2,\muc^2)\otimes K_{kl}\left( \smallfrac{m_c^2}{\muc^2}\right)\otimes\bigGamma^{\nthree}_{lj}(\muc^2,Q^2) ,   \label{eq:transmatrix}
\end{equation}
and we have used the fact that the evolution matrices can be inverted
by evolving backwards: $\Gamma_{ij}(\muc^2,Q^2)$ is the inverse of
$\Gamma_{ij}(Q^2,\muc^2)$. Note that while $K_{ij} (m_c^2/\muc^2)$
contains no large logarithms (since $\muc^2 \sim m_c^2$), the large
logarithms of $Q^2/\muc^2$ resummed in the evolution factors are
mismatched, so $T_{ij} (Q^2,\muc^2,m_c^2)$ also resums large
logarithms.
If the evolution factors are expanded to any given fixed order in $\as$,
the $\muc$ dependence of $T_{ij}$ disappears and $T_{ij}(Q^2,\muc^2,m_c^2)=K_{ij}(m_c^2/Q^2)$.

Having established the relation between the PDFs in the two schemes, Eq.~\eqref{eq:transpdf},
it is now interesting to use it to write the massive 3FS result Eq.~\eqref{eq:Fthree}
in terms of the 4FS PDFs $f_i^{\nfour}(Q^2)$ (and also in terms of $\alpha_s^{\nfour}(Q^2)$ through Eq.~\eqref{eq:matchalpha},
though we leave the dependence implicit).
This will be needed for the FONLL construction described in the next section.
Substituting the inverse of the transformation Eq.~\eqref{eq:transpdf} into Eq.~\eqref{eq:Fthree}, we get
\bea
  F^{\nthree} (Q^2,m_c^2) 
&= \sum_{i,j = g, q, \bar{q}, c, \bar{c}}
  C_i^{\nthree} \left(\smallfrac{m_c^2}{Q^2}\right)\otimes
  T_{ij}^{-1} (Q^2,\muc^2,m_c^2)\otimes
  f_j^{\nfour} (Q^2), \label{eq:Fthreebar}
\eea
where
\begin{equation}
  T_{ij}^{-1} (Q^2,\muc^2,m_c^2)  =
  \sum_{k,l = g, q, \bar{q}, c, \bar{c}}\bigGamma_{ik}^{\nthree}(Q^2,\muc^2)\otimes
  K_{kl}^{-1}\left( \smallfrac{m_c^2}{\muc^2}\right)\otimes\Gamma^{\nfour}_{lj}(\muc^2,Q^2)
   ,   \label{eq:transmatrixinv}
\end{equation}
and $K_{ij}^{-1}$ is obtained from Eq.~(\ref{eq:matchpdfexp4}) by inverting term by term.
The large logarithms of $m_c^2/Q^2$ in $C^{\nthree}_i$ must then cancel
term by term with corresponding large logarithms in $T^{-1}_{ij}$,
resulting from the mismatch of the two evolution factors
$\Gamma^{\nfour}_{ij}$ and $\bigGamma^{\nthree}_{ij}$: in other words $T^{-1}_{ij}$
provides the correct subtraction terms for $C^{\nthree}_i$. Moreover
$T^{-1}_{ij}$ also gives automatically the correct finite parts of the
subtraction.

Comparing Eq.~(\ref{eq:Fthreebar}) with Eq.~(\ref{eq:Ffour}), and noting
that to a given order in resummed perturbation theory the structure
function (being physical) must be independent of the scheme, we see
immediately that
\begin{equation}
C_i^{\nfour} \left(\smallfrac{m_c^2}{Q^2}\right) = 
\sum_{j = g, q, \bar{q}, c, \bar{c}}
  C_j^{\nthree} \left( \smallfrac{m_c^2}{Q^2}\right)\otimes
  T_{ji}^{-1} (Q^2,\muc^2,m_c^2) .\label{eq:coeffthreesub} 
\end{equation}
Note that once the coefficient function is expanded out to fixed order 
in $\alpha_s$, there is nothing to prevent us from setting $\muc^2=Q^2$ in
Eq.~(\ref{eq:coeffthreesub}): this simplifies the expressions by
setting both evolution factors to unity, so that
\begin{equation}
C_i^{\nfour} \left(\smallfrac{m_c^2}{Q^2}\right) = 
\sum_{j = g, q, \bar{q}, c, \bar{c}}
  C_j^{\nthree} \left(\smallfrac{m_c^2}{Q^2}\right)\otimes
  K_{ji}^{-1} \left( \smallfrac{m_c^2}{Q^2}\right) ,\label{eq:C4C3K} 
\end{equation}
the large logarithms of $Q^2/m_c^2$ now being those in $K^{-1}_{ij}$. When truncated to  
any given fixed order in perturbation theory,
Eq.~(\ref{eq:coeffthreesub}) and Eq.~(\ref{eq:C4C3K}) will yield
identical results for the coefficient functions $C_i^{\nfour}$, independent of the matching scale $\mu_c$. 

It is interesting to observe that the massive 4FS result
Eq.~\eqref{eq:Ffour}, introduced originally as the result of a
collinear factorization with massive quarks, has now been derived from
the massive 3FS result Eq.~\eqref{eq:Fthree} after scheme change, Eq.~\eqref{eq:C4C3K},
which removes all its collinear logarithms. The matching condition
Eq.~(\ref{eq:matchdef}) then ensures that in the massless limit 
\beq
\label{eq:limit} \lim_{m_c\to0}
C_i^{\nfour}\left(\smallfrac{m_c^2}{Q^2}\right) = C_i^{\nfour}(0),
\eeq
as expected from Eq.~\eqref{eq:C4limit}.  For this to work
properly it is essential that the coefficient function $C_i^{\nthree}$
and matching matrix $K_{ij}$ are always evaluated to the same fixed
order: if the coefficient function has terms of higher order than the matching
matrix there will be uncancelled logarithms, while if the matching
matrix has terms of higher order than the coefficient function it will
be trying to cancel logarithms which are not there.
This observation is of great importance if one wishes to combine
the results obtained in different schemes, as done in the FONLL prescription.

We have thus shown that starting from the massive 3FS result
Eq.~\eqref{eq:Fthree}, and re-expressing it in terms of 4FS PDFs,
we obtain a result equivalent to the massive 4FS result
Eq.~\eqref{eq:Ffour}: in other words we can use the resummation of the
massless collinear logarithms performed in the massless 4FS by the
evolution Eq.~\eqref{eq:Efour} to resum the large logarithms in the
massive 3FS. The result is at the heart of the ACOT scheme
\cite{Aivazis:1993kh,Aivazis:1993pi}: formally order by order in
perturbation theory
\begin{align}
  F_{\rm ACOT} (Q^2,m_c^2) &= \sum_{i = g, q, \bar{q}, c, \bar{c}}
  C_i^{\nfour} \left(\smallfrac{m_c^2}{Q^2} \right)\otimes
  f_i^{\nfour} (Q^2), \nonumber\\
 &= \sum_{i,j = g, q, \bar{q}, c, \bar{c}}
  C_i^{\nthree} \left(\smallfrac{m_c^2}{Q^2} \right)\otimes
  K_{ij}^{-1} \left( \smallfrac{m_c^2}{Q^2}\right)\otimes
  f_j^{\nfour} (Q^2). \label{eq:ACOT}
\end{align}
The form of Eq.~(\ref{eq:ACOT}) is interesting: it combines the PDFs
$f_i^{\nfour}$ evolved in the 4FS with the coefficient
function $C_i^{\nthree}$ computed in the massive 3FS, the matching
conditions $K_{ij}$ linking the two schemes subtracting the unresummed
collinear logarithms from the massive coefficient functions so that
the coefficients convoluted with the PDFs are free from collinear
logarithms and thus have a well behaved perturbative expansion.

%% file: sections/sec-fonll.tex
\section{Combining Fixed Order and Resummation}
\label{sec:fonll}

In the previous Section we showed that a consistent scheme change
relates a 3FS calculation, which does not factorizes the
collinear logarithms due to the charm quark, to a 4FS
calculation, in which the collinear logarithms are resummed and the
massive effects are included into the 4FS coefficient
function. An alternative way of combining massive coefficient
functions in the 3FS with massless coefficient functions
in the 4FS is the FONLL construction
\cite{Buza:1996wv,Cacciari:1998it}. 

In most applications so far FONLL
has been used with the assumption that charm is generated entirely
perturbatively (so there is no `intrinsic' charm): the structure
function in the 3FS can then be expressed entirely in
terms of light partons. We thus require a precise all-order definition
of what we mean by `zero intrinsic charm' before we can obtain
definite all-order results.

Here we will summarize the main features of the FONLL construction in
Sect.~\ref{sec:fonll-con}, and explain its relation to ACOT in
Sect.~\ref{sec:ACOT}. We then discuss the definition of intrinsic charm,
and the transition from 3FS PDFs to 4FS PDFs
in Sect.~\ref{sec:ECIC}, and present FONLL results for structure
functions without intrinsic charm in Sect.~\ref{sec:fonll-zic}
(corresponding to the NNLO results in \cite{Forte:2010ta}, now
formally generalized to all orders), and the corrections necessary
when intrinsic charm is included in Sect.~\ref{sec:fonll-ic}. We then
go on to show the relation of the FONLL scheme without intrinsic charm
and the S-ACOT schemes in Sect.~\ref{sec:SACOT}.
We finally discuss in Sect.~\ref{sec:damping}
a phenomenological damping factor included in the original FONLL formulation.

\subsection{The FONLL construction}
\label{sec:fonll-con}
 
The general FONLL construction \cite{Buza:1996wv,Cacciari:1998it} is
based on the observation that to evaluate cross-sections consistently
both in the threshold and the high energy region, it is sufficient to
any given order in perturbation theory to add the massive 3FS result
(which includes all charm mass effects to fixed order) to the massless
4FS result (which performs the resummation of all large logarithms at
high energy), and then subtract any doubly counted contributions.

The FONLL construction only involves physical quantities computed in
well-defined (massive 3F or massless 4F) factorization schemes, and thus
side steps issues related to the existence of more novel factorization
schemes. In particular this means that in FONLL it is straightforward
to write down expressions at any order in perturbation theory: all one
has to do is evaluate the relevant massive diagrams in the massive
scheme, and combine them linearly with the corresponding massless
calculations. The only nontrivial part is then to identify the double
counting.

Structure functions calculated with four flavours in the FONLL method are thus given by
\begin{equation}
F_{\rm FONLL} \left(Q^2,m_c^2 \right)
= F^{\nfour} \left(Q^2,0 \right)
+ F^{\nthree}\left( Q^2,m_c^2 \right)
- \textrm{d.c.} \ .
\end{equation} 
The double counting term can be obtained as the massless limit of the
massive 3FS result, and corresponds to the fixed-order expansion of the
massless 4FS result.  The massless limit of the massive coefficient
functions is however divergent, due to the presence of unsubtracted
massive collinear logarithms.  A proper definition of this term is
given by
\begin{equation}
  F^{\nthreez} (Q^2,m_c^2) 
=  \sum_{i = g, q, \bar{q}, c ,\bar{c}}
  C_i^{\nthreez} \left( \smallfrac{m_c^2}{Q^2} \right)\otimes
  f_i^{\nthree} (Q^2), \label{eq:Fthreezero}
\end{equation}
where we used Eq.~\eqref{eq:masslimdef} to define 
the (singular) massless limit $C^\nthreez_i ( m_c^2/Q^2)$ of the massive coefficient functions $C_i^{\nthree}( m_c^2/Q^2)$.
In this limit all terms which vanish as $m_c^2\to 0$ are removed, and all that remains are the finite terms and collinear logarithms,
as explicitly shown in Eq.~\eqref{eq:matchdef1}.

The structure functions in the FONLL prescription are thus given by
\begin{equation}
F_{\rm FONLL} \left(Q^2,m_c^2 \right)
= F^{\nfour} \left(Q^2,0 \right)
+ \left[F^{\nthree}\left( Q^2,m_c^2 \right)
-F^{\nthreez}\left(Q^2,m_c^2\right)\right] .
\label{eq:FONLL}
\end{equation} 
Clearly Eq.~(\ref{eq:FONLL}) does what we want it to: in particular
when $Q^2\gg m_c^2$ the terms in square brackets vanish as a power of
$m_c^2/Q^2$, and we recover the massless coefficient function in the 4FS.
Likewise, when $Q^2\sim m_c^2$, we can write
\begin{align}
F_{\rm FONLL} \left(Q^2,m_c^2 \right)
&= F^{\nthree}\left( Q^2,m_c^2 \right)+ \left[ F^{\nfour} \left(Q^2,0 \right)
-F^{\nthreez}\left(Q^2,m_c^2\right)\right]\nonumber\\
&\equiv F^{\nthree}\left( Q^2,m_c^2 \right)+ F^{(d)} \left(Q^2,m_c^2 \right).
\label{eq:FONLLd}
\end{align} 
where in Ref.~\cite{Forte:2010ta} the term in square brackets is
referred to as the `difference term' $F^{(d)}$. While this term is nonzero for
$Q^2\sim m_c^2$, it is subleading in $\alpha_s(Q^2)$, since when
$Q^2\sim m_c^2$ there are no large logarithms. Thus
Eq.~(\ref{eq:FONLL}) gives a structure function which is correct at
high energy, up to power suppressed corrections, and correct in the
threshold region up to subleading corrections.

\subsection{Comparison to ACOT}
\label{sec:ACOT}

One way of using the FONLL construction would be simply to compute the
three ingredients $F^{\nthree}$, $F^{\nthreez}$ and $F^{\nfour}$, using the
factorized expression in the 3-flavour and 4-flavour schemes,
Eqs.~\eqref{eq:Fthree}, \eqref{eq:Fthreezero}, \eqref{eq:Fzero}, and combine
them linearly according to Eq.~\eqref{eq:FONLL}. In practice this is
awkward, because it means one has to work simultaneously with PDFs in
two different schemes. Thus instead it is more convenient to use the matching
of the two schemes, Eq.~(\ref{eq:matchpdf}), to write
Eq.~(\ref{eq:FONLL}) in terms of PDFs in the 4FS and thus
in the form Eq.~\eqref{eq:Ffour}. In this way we can identify explicit
expressions for the mass dependent coefficient functions
$C_i^{\nfour}(m_c^2/Q^2)$.

We showed in Sect.~\ref{sec:matching} that using the scheme change we can write $F^{\nthree}$ in the form
\begin{equation}
  F^{\nthree} (Q^2,m_c^2) = \sum_{i,j = g, q, \bar{q}, c, \bar{c}}
  C_i^{\nthree} \left(\smallfrac{m_c^2}{Q^2} \right)\otimes
  K_{ij}^{-1} \left( \smallfrac{m_c^2}{Q^2}\right)\otimes
  f_j^{\nfour} (Q^2). \label{eq:Fthreefour}
\end{equation}
Consistently, in the massless limit, we have
\begin{equation}
  F^{\nthreez}  (Q^2,m_c^2) = \sum_{i,j = g, q, \bar{q}, c, \bar{c}}
  C_i^{\nthreez} \left(\smallfrac{m_c^2}{Q^2} \right)\otimes
  K_{ij}^{-1} \left( \smallfrac{m_c^2}{Q^2}\right)\otimes
  f_j^{\nfour} (Q^2). \label{eq:Fthreefourzero}
\end{equation}
Substituting into Eq.~(\ref{eq:FONLL}), we have 
\begin{align}
F_{\rm FONLL} \left(Q^2,m_c^2 \right) &=  \sum_{i,j = g, q, \bar{q}, c, \bar{c}}
  \bigg[C_i^{\nthree} \left( \smallfrac{m_c^2}{Q^2}\right)-C_i^{\nthreez} \left( \smallfrac{m_c^2}{Q^2}\right)\bigg]\otimes
  K^{-1}_{ij}\left( \smallfrac{m_c^2}{Q^2}\right)\otimes f_j^{\nfour} (Q^2)\nonumber\\
  &\qquad + \sum_{i = g, q, \bar{q}, c, \bar{c}}
  C_i^{\nfour}(0)\otimes f_i^{\nfour} (Q^2).
 \label{eq:FONLL4}
\end{align}
Now however we find an interesting simplification: the difference term vanishes identically,
since due to the matching condition Eq.~(\ref{eq:matchpdf}) the terms with coefficient functions in the
4FS precisely cancel those from the massless limit of the 3FS, Eq.~\eqref{eq:masslimdef}.
Eq.~(\ref{eq:FONLL4}) can thus be written simply as
\bea
  F_{\rm FONLL} (Q^2,m_c^2) &= \sum_{i,j = g, q, \bar{q}, c, \bar{c}}
  C_i^{\nthree} \left(\smallfrac{m_c^2}{Q^2} \right)\otimes
  K_{ij}^{-1} \left( \smallfrac{m_c^2}{Q^2}\right)\otimes
  f_j^{\nfour} (Q^2) \nonumber\\
&= \sum_{i = g, q, \bar{q}, c, \bar{c}}
  C_i^{\nfour} \left(\smallfrac{m_c^2}{Q^2} \right)\otimes
  f_i^{\nfour} (Q^2) \nonumber\\
&= F_{\rm ACOT} \left(Q^2,m_c^2 \right),
  \label{eq:ACOT2}
\eea
with no massless coefficient functions at all.  This shows that when we make no theoretical 
assumption about the 4FS PDFs at the initial scale, the FONLL construction of the structure function is equivalent to ACOT order by order in perturbation theory. Indeed, the above 
manipulations can be viewed as an alternative
all-order derivation of the ACOT result using the FONLL
construction. The only essential ingredients are the existence of the
3-flavour and 4-flavour factorization schemes, and the matching relations
which allow them to be related together.

In retrospect this result should not have been too surprising. When
$Q^2\gg m_c^2$, the massless 4FS PDFs $f_i^{\nfour}(Q^2)$
$i,j = g, q, \bar{q}, c, \bar{c}$ resum all large logarithms through
solution of the evolution equations Eq.~\eqref{eq:Efour}. It follows, as
explained after Eq.~\eqref{eq:C4C3K}, that the coefficient functions
convoluted with these PDFs, which contain all dependence on $m_c^2$,
must be free of large logarithms, and thus have a well behaved
perturbative expansion in $\alpha_s^{\nfour}(Q^2)$. Since the
structure function is a physical quantity, the coefficients in the
perturbative expansion of the coefficient function must (for a given
factorization and renomalization scheme) be then unique to each order,
which is indeed what we find.

It is important to note however that in deriving this result we made
no assumption about the origin of charm, and in particular we did not
assume that charm is generated purely perturbatively. To make contact
with other work on FONLL, in particular Ref.~\cite{Forte:2010ta}, where
this assumption is an integral part of the construction, we first need
to define carefully what we mean when we assume that all charm is
generated perturbatively, i.e.\ when there is no `intrinsic'
charm. This will tell us how to modify the expressions given in
Ref.~\cite{Forte:2010ta} to incorporate intrinsic charm, and in turn
help us to understand better the role of intrinsic charm in the
formulation of ACOT.

\subsection{Intrinsic Charm}
\label{sec:ECIC}

In the previous sections we made no attempt to distinguish between
extrinsic (perturbative) and intrinsic charm: we have been agnostic about the nature of the initial
charm distributions at $Q_0$, in either 3- or 4-flavour schemes,
which can be fitted at that scale and then perturbatively evolved to the scale $Q$.
However in the more conventional formalisms 
there are no fitted charm PDFs:
zero `intrinsic' charm is an implicit part of the construction. The
definition of intrinsic charm is in truth rather ambiguous, and any
condition of zero intrinsic charm must reflect this ambiguity:
conditions can be made in different schemes and at different scales,
and will in general all differ by subleading terms. Thus the only
formalism devoid of such ambiguities is the complete 4-flavour
formalism adopted above, where the charm PDF takes part to DGLAP evolution
and it is fitted together with the light flavour PDFs.
Nevertheless it is interesting to consider
the case of zero intrinsic charm, appropriately defined, in order to
make contact with previous work, in particular Ref.~\cite{Forte:2010ta}.

We first note that due to factorization all the information about the
nature of the target hadron, and in particular whether or not it
contains intrinsic charm, is contained in its PDFs. Since in a
particular renormalization and factorization scheme the PDF evolution
is also target independent, the only place where intrinsic charm can
enter is in the boundary conditions for the perturbative evolution of
the PDFs.

This necessarily implies that once the massive coefficient functions
$C_i^{\nfour}(m_c^2/Q^2)$ have been correctly computed to a given
order for calculations with intrinsic charm, the very same coefficient
functions must also hold to the same order when there is no intrinsic
charm. The same is true of the matching matrix $K_{ij}(m_c^2/Q^2)$,
since this can be defined entirely in terms of coefficient
functions. This is a straightforward consequence of factorization:
coefficient functions are by construction hard cross sections, and are
thus independent of the target hadron.  Unfortunately however the
converse is not true: coefficient functions computed to a given order
for the special case of no intrinsic charm might need correcting in
the more general case when intrinsic charm is included. We shall see
below that this is indeed the case.

A naive definition of zero intrinsic charm would be that it vanish in
the 4-flavour scheme at the initial scale:
$f_c^{\nfour} (x, Q_0^2)= f_{\bar{c}}^{\nfour} (x, Q_0^2) = 0$.
Unfortunately this definition is rather ambiguous: instead of the
rather arbitrary starting scale $Q_0$, one might instead prefer the
scale of the charm mass $m_c$, or indeed the threshold scale
$W=2m_c$. However once the scale is chosen, this would mean that
in the 4FS all charm is `extrinsic', i.e.\ it is generated
dynamically by perturbative evolution, so
$f_c^{\nfour} (x, Q^2)$ and $f_{\bar{c}}^{\nfour} (x, Q^2)$ can be
expressed entirely in terms of light quark PDFs.

A much better characterization of intrinsic charm is to define it as
the charm PDF in the 3FS, where the charm PDF does not evolve, and
thus, one could argue, there is no extrinsic charm. Thus, zero
intrinsic charm means that
\begin{equation} 
f_c^{\nthree} = f_{\bar{c}}^{\nthree} = 0. \label{eq:zicthree}
\end{equation}
This is a very natural assumption to make, since in the massive 3FS
scheme the charm PDFs are scale independent, so there is no ambiguity
about scale choice. Unfortunately it means that there is in general no
scale at which the 4FS charm PDFs vanish, as can be seen from the
matching condition Eq.~(\ref{eq:matchpdf}): if charm in one scheme is
zero, in the other it will be generally nonzero already at
$\Ord(\alpha_s)$ if $\mu\neq m_c$, due to the off-diagonal term
$K_{cg}$ in the matching condition, and at $\Ord(\alpha_s^2)$ even for
$\mu = m_c$ due to the presence of non-logarithmic terms at this order.

The condition Eq.~(\ref{eq:zicthree}) can however always be turned into
a nonzero boundary condition for perturbative massless evolution. To
see how this works, we first write out the matching conditions
Eq.~(\ref{eq:matchpdf}) separating out the light partons from the charm
partons:
\begin{align}  
f_i^{\nfour} (\muc^2)  &= \sum_{j = g, q, \bar{q}} K_{ij}\left( \smallfrac{m_c^2}{\muc^2}\right)\otimes f_j^{\nthree} (\muc^2) + \sum_{j = c,\bar{c}} K_{ij}\left( \smallfrac{m_c^2}{\muc^2}\right)\otimes f_j^{\nthree},\qquad i = g,q, \bar{q},
\label{eq:matchpdfq}
\\
f_i^{\nfour} (\muc^2)  &= \sum_{j = g, q, \bar{q}} K_{ij}\left( \smallfrac{m_c^2}{\muc^2}\right)\otimes f_j^{\nthree} (\muc^2) + \sum_{j = c,\bar{c}} K_{ij}\left( \smallfrac{m_c^2}{\muc^2}\right)\otimes f_j^{\nthree},\qquad i = c, \bar{c}.\label{eq:matchpdfc}
\end{align} 
When there is no intrinsic charm, Eq.~(\ref{eq:zicthree}), the second
term in each of these equations vanishes, and in particular
Eq.~(\ref{eq:matchpdfc}) becomes simply
\begin{equation}  
f_c^{\nfour} (\muc^2)  = \sum_{j = g, q, \bar{q}} K_{cj}\left( \smallfrac{m_c^2}{\muc^2}\right)\otimes f_j^{\nthree} (\muc^2), \qquad \(f_{c,\bar c}^{\nthree} = 0\).\label{eq:fcfourthreezic}
\end{equation}
Thus when there is no intrinsic charm case, it is possible (and often
convenient) to invert Eq.~\eqref{eq:matchpdfq} to express the light
3FS PDFs in terms of \emph{only light} 4FS PDFs, as
\beq 
f_i^{\nthree}
(\muc^2) = \sum_{j = g, q, \bar{q}} \tilde{K}_{ij}^{-1}\left(
  \smallfrac{m_c^2}{\muc^2}\right)\otimes f_j^{\nfour} (\muc^2), \qquad i =
q, \bar{q}, g, \qquad
\(f_{c,\bar c}^{\nthree} = 0\),\label{eq:fthreefourzic}
\eeq
where $\tilde{K}_{ij}$ is the matching matrix
restricted to the subspace of light partons, $i,j = g, q, \bar{q}$, so
that the inverse is taken in this subspace. Substituting into
Eq.~\eqref{eq:fcfourthreezic} we find, for $Q_0 \sim m_c$
\begin{equation}  
f_c^{\nfour} (Q_0^2) = \sum_{j,k = g, q, \bar{q}} K_{cj}\left( \smallfrac{m_c^2}{Q_0^2}\right)
\otimes\tilde{K}^{-1}_{jk} \left( \smallfrac{m_c^2}{Q_0^2}\right)
\otimes f_k^{\nfour} (Q_0^2),\qquad \(f_{c,\bar c}^{\nthree} = 0\)
 ,\label{eq:zicbc}
\end{equation}
which is the required boundary condition expressing the charm PDF in
terms of the light PDFs at the starting scale. Note that if we choose
$Q_0 = m_c$, $f_c^{\nfour} (Q_0^2)$ will be $O(\alpha_s^2)$ and thus
presumably very small: still, it is always nonzero in general. However
all charm, at any scale, is still determined perturbatively from the
light parton PDFs, and is thus extrinsic. The condition
Eq.~\eqref{eq:zicbc} may be consistently applied at any scale
$Q_0\sim m_c$: changes in $Q_0$ only introduce subleading
corrections. However it will not hold for $Q_0\gg m_c$, since then
these formally subleading corrections will be accompanied by large
logarithms.

\subsection{FONLL with zero intrinsic charm}
\label{sec:fonll-zic}

Now that we have a definition of intrinsic charm, we can apply the
FONLL construction under the assumption that all charm is generated
perturbatively \cite{Forte:2010ta}. The treatment given here will be
entirely explicit, to any order in perturbation theory.

When there is no intrinsic charm, it is possible to write the structure
function in the massive scheme, $F^{\nthree}$, entirely in terms of
the light partons in the 4FS: combining
Eq.~(\ref{eq:Fthree}) and Eq.~(\ref{eq:fthreefourzic}),
\begin{align}
  F^{\nthree} (Q^2,m_c^2)\Big|_{\rm zic} &=  \sum_{i = g, q, \bar{q}}
  C_i^{\nthree} \left( \smallfrac{m_c^2}{Q^2}\right)\otimes
  f_i^{\nthree} (Q^2)\nonumber\\
&=  \sum_{i,j = g, q, \bar{q}}
  C_i^{\nthree} \left( \smallfrac{m_c^2}{Q^2}\right)\otimes
  \tilde{K}^{-1}_{ij}\left( \smallfrac{m_c^2}{Q^2}\right)\otimes f_j^{\nfour} (Q^2) , 
\label{eq:FthreezIC}
\end{align}
where the subscript `zic' stands for `zero intrinsic charm'. Thus when
there is no intrinsic charm the coefficient functions $C_{c,\bar c}^{\nthree}$
are not needed, and likewise the matching terms $K_{ic}$, $K_{i\bar c}$. Of course
these terms are still nonzero, but when Eq.~(\ref{eq:zicthree}) holds
they are no longer needed for the evaluation of $F^{\nthree}$, and can
thus be ignored. Like Eq.~(\ref{eq:Fthree}), this expression only holds
for $Q^2 \sim m_c^2$: although the light PDFs are in the 4FS,
and thus resum collinear logarithms in the light sector,
Eq.~(\ref{eq:fthreefourzic}) is fixed order, so the large logarithms of
$m_c^2/Q^2$ in the heavy quark sector are not resummed.

In Ref.~\cite{Forte:2010ta} Eq.~(\ref{eq:FthreezIC}) is written as
\begin{equation}
  F^{\nthree}(Q^2,m_c^2)\Big|_{\rm zic} =  \sum_{i = g, q, \bar{q}}
  B^{\nfour}_i \left( \smallfrac{m_c^2}{Q^2}\right)\otimes f_i^{\nfour} (Q^2) ,
\label{eq:FthreezICB}
\end{equation}
having defined
\begin{equation}
B^{\nfour}_i\left(\smallfrac{m_c^2}{Q^2} \right) = 
\sum_{j = g, q, \bar{q}}
  C_j^{\nthree} \left(\smallfrac{m_c^2}{Q^2} \right)\otimes
  \tilde{K}_{ji}^{-1} \(\smallfrac{m_c^2}{Q^2}\) ,\label{eq:Bdef} 
\end{equation}
with inverse
\begin{equation}
C_i^{\nthree}\left(\smallfrac{m_c^2}{Q^2} \right) = 
\sum_{j = g, q, \bar{q}}
  B^{\nfour}_j \left(\smallfrac{m_c^2}{Q^2} \right)\otimes
  K_{ji} \(\smallfrac{m_c^2}{Q^2}\), \qquad i=g, q, \bar q
 .\label{eq:Bdefinv} 
\end{equation}
The similarity with the more general ACOT expression
Eq.~(\ref{eq:ACOT}) is obvious: the only difference in fact is that the
sum extends only over the light partons, and the inverse of the
matching matrix, $\tilde{K}^{-1}$, is likewise taken in the light
parton subspace.

\begin{figure}[t]
  \centering
  \includegraphics[width=0.8\textwidth]{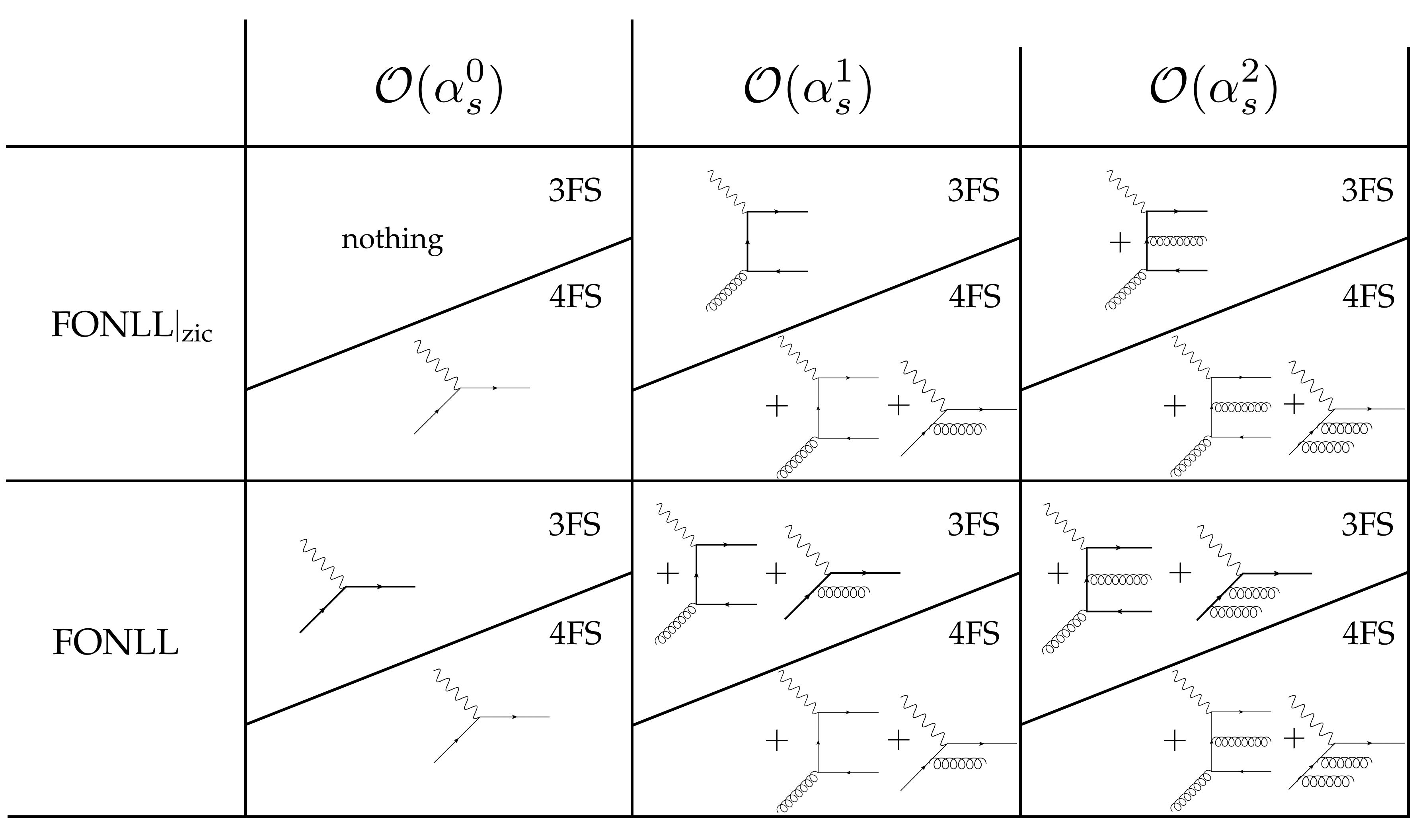}
  \caption{Representative diagrams included at each order in $\as$ in the massive 3FS and massless 4FS components
    entering the FONLL formula Eq.~\eqref{eq:FONLL}, in the simpler case of no intrinsic charm 
    (upper line, corresponding to Eq.~\eqref{eq:FONLLzIC})
    and in the full case (lower line, corresponding to Eq.~\eqref{eq:FONLLsimp}).
    All quark lines represent the charm quark: bold lines means that mass effects are retained in the coefficient function,
    while light lines means that the charm mass has been set to zero.
    Diagrams with corrections to the gluon propagator through a virtual heavy quark loop are not shown.}
  \label{fig:counting}
\end{figure}

Substituting Eq.~(\ref{eq:FthreezICB}) in the general expression
Eq.~(\ref{eq:FONLL}), and Eq.~(\ref{eq:Fzero}) for the massless term,
we immediately obtain the FONLL expression for the structure function
when there is no intrinsic charm:
\begin{align}
F_{\rm FONLL}(Q^2,m_c^2 )\Big|_{\rm zic}
  &=  \sum_{i = g, q, \bar{q}}
    \[B^{\nfour}_i \left( \smallfrac{m_c^2}{Q^2}\right)-B^{\nfourz}_i \left( \smallfrac{m_c^2}{Q^2}\right)+C_i^{\nfour}(0)\]
    \otimes f_i^{\nfour} (Q^2)\nonumber\\
  &\qquad\qquad  + \sum_{i = c, \bar{c}}
    C_i^{\nfour}(0)\otimes f_i^{\nfour} (Q^2).
    \label{eq:FONLLzIC}
\end{align}
Here, in analogy to  Eq.~(\ref{eq:Bdef}),
\bea
B^{\nfourz}_i\left(\smallfrac{m_c^2}{Q^2} \right) &= 
\sum_{j = g, q, \bar{q}}
  C_j^{\nthreez} \left(\smallfrac{m_c^2}{Q^2}\right)\otimes
  \tilde{K}_{ji}^{-1} \(\smallfrac{m_c^2}{Q^2}\) \nonumber\\
&= \sum_{j = g, q, \bar{q}}\sum_{k = g, q, \bar{q}, c, \bar c}
  C_k^{\nfour}(0) \otimes K_{kj} \(\smallfrac{m_c^2}{Q^2}\) \otimes
  \tilde{K}_{ji}^{-1} \(\smallfrac{m_c^2}{Q^2}\) \nonumber\\
&= C_i^{\nfour}(0) + \sum_{k = c, \bar c} \sum_{j = g, q, \bar{q}}
   C_k^{\nfour}(0)\otimes K_{kj} \(\smallfrac{m_c^2}{Q^2}\) \otimes
  \tilde{K}_{ji}^{-1} \(\smallfrac{m_c^2}{Q^2}\) ,\label{eq:Bdef0} 
\eea
where we have used Eq.~\eqref{eq:masslimdef} in the second step, and
Eq.~\eqref{eq:Ktilinvgg} (see Appendix \ref{sec-appendixmat}) in the last. For
$Q^2\gg m_c^2$, the first two terms in Eq.~\eqref{eq:FONLLzIC} are
manifestly of order $m_c^2/Q^2$, and thus the massless limit
Eq.~\eqref{eq:Fzero} is recovered.
The diagrams contributing to the FONLL result in the zero intrinsic charm case
are shown schematically in Fig.~\ref{fig:counting} (upper row).

Substituting the last line into Eq.~(\ref{eq:FONLLzIC}) gives the alternative expression
\begin{align}
&F_{\rm FONLL}(Q^2,m_c^2)\Big|_{\rm zic} =  \sum_{i = g, q, \bar{q}}
  B^{\nfour}_i \left( \smallfrac{m_c^2}{Q^2}\right)\otimes f_i^{\nfour} (Q^2)
+ \sum_{i = c, \bar{c}}
  C_i^{\nfour}(0)\otimes f_i^{\nfour} (Q^2)
\nonumber\\
  &\qquad  - \sum_{k = c, \bar c} \sum_{i,j = g, q, \bar{q}}
   C_k^{\nfour}(0)\otimes K_{kj} \(\smallfrac{m_c^2}{Q^2}\) \otimes
  \tilde{K}_{ji}^{-1} \(\smallfrac{m_c^2}{Q^2}\)\otimes f_i^{\nfour}(Q^2).
 \label{eq:FONLLzIC1}
\end{align}
The last two terms are the difference term Eq.~\eqref{eq:FONLLd}. At
the initial scale $Q_0$ the difference term is thus precisely zero,
due to the boundary condition Eq.~\eqref{eq:zicbc} used when there is
no intrinsic charm. For $Q> Q_0$, since the 4FS PDFs
are constrained to evolve using 4FS evolution
Eq.~(\ref{eq:Efour}), generating charm perturbatively, they will
contain higher order logarithms which do not cancel: with the matching
matrix truncated to order $\alpha_s^p$,
\begin{equation}
 f_i^{\nfour}(Q^2)  - \sum_{j,k = g, q, \bar{q}} 
 K_{ij}\left( \smallfrac{m_c^2}{Q^2}\right)\otimes \tilde{K}^{-1}_{jk}\left( \smallfrac{m_c^2}{Q^2}\right)\otimes f_k^{\nfour}(Q^2)
 \sim \Ord\(\alpha_s^{p+1}\log^{p+1}\smallfrac{m_c^2}{Q^2}\) .\label{eq:fthreecord}
\end{equation}
It follows that whenever $Q\sim m_c$, so that
$\log(m_c^2/Q^2)$ is not too large, the difference term
is always subleading, as required.

\subsection{FONLL including intrinsic charm}
\label{sec:fonll-ic}

When we drop the assumption that the intrinsic charm is zero, we can
go through the same argument as in the previous Section expressing
$F^{\nthree}$ in terms of $f^{\nfour}_i$, but keeping the nonzero
$f_{c,\bar{c}}^{\nthree}$ terms: we then find
\bea
F^{\nthree} (Q^2,m_c^2)
&= \sum_{i = g, q, \bar{q}} C_i^{\nthree}\left(\smallfrac{m_c^2}{Q^2} \right)\otimes f_i^{\nthree}(Q^2)
+ \sum_{i = c,\bar c} C_i^{\nthree}\left(\smallfrac{m_c^2}{Q^2} \right) \otimes  f_i^{\nthree} \nonumber\\
&= \sum_{i,k = g, q, \bar{q}} B^{\nfour}_k\left(\smallfrac{m_c^2}{Q^2} \right)\otimes K_{ki}\left(\smallfrac{m_c^2}{Q^2} \right)\otimes f_i^{\nthree}(Q^2)
+ \sum_{i = c,\bar c} C_i^{\nthree}\left(\smallfrac{m_c^2}{Q^2} \right) \otimes  f_i^{\nthree} \nonumber\\
&= \sum_{k = g, q, \bar{q}} B^{\nfour}_k\left(\smallfrac{m_c^2}{Q^2} \right)\otimes f_k^{\nfour}(Q^2)\nonumber\\
&\qquad + \sum_{i = c,\bar c} \bigg[C_i^{\nthree}\left(\smallfrac{m_c^2}{Q^2} \right) - \sum_{k = g, q, \bar{q}} B^{\nfour}_k\left(\smallfrac{m_c^2}{Q^2} \right) \otimes K_{ki}\left(\smallfrac{m_c^2}{Q^2} \right) \bigg] \otimes  f_i^{\nthree},
\label{eq:F3derivation}
\eea
where in the second line we used Eq.~\eqref{eq:Bdefinv}, and in the third Eq.~(\ref{eq:matchpdfq}).
We recognize in the first term the zero intrinsic charm result, Eq.~\eqref{eq:FthreezICB}, so 
defining
\begin{equation}
F^{\nthree} (Q^2,m_c^2)=F^{\nthree} (Q^2,m_c^2)\Big|_{\rm zic}+\Delta F^{\nthree} (Q^2,m_c^2) ,
\label{eq:FthreeICsep}
\end{equation}
the intrinsic charm contribution to the structure function in the massive 3FS scheme is 
\begin{equation}
\Delta F^{\nthree} (Q^2,m_c^2)=\sum_{i = c, \bar{c}}\bigg[
  C_i^{\nthree} \left( \smallfrac{m_c^2}{Q^2}\right)
 - \sum_{j = g, q, \bar{q}} 
B^{\nfour}_j \left( \smallfrac{m_c^2}{Q^2}\right)
\otimes K_{ji}\left( \smallfrac{m_c^2}{Q^2}\right)\bigg]\otimes
f_i^{\nthree}  .\label{eq:FthreeIC}
\end{equation}
Note that this precise form derives from having used the coefficient functions $B^\nfour$
for writing the zero intrinsic charm contribution.

We can no longer use Eq.~\eqref{eq:fthreefourzic} to express
$f_{c,\bar{c}}^{\nthree}$ in terms of 4FS PDFs: instead we have
to use the inverse of Eq.~\eqref{eq:matchpdf}: for $i=c,\bar{c}$
\bea
f_i^{\nthree} &= \sum_{j = g,q,\bar q, c, \bar{c}} K^{-1}_{ij}\left( \smallfrac{m_c^2}{Q^2}\right) \otimes f_j^{\nfour}(Q^2)\nonumber\\
& =\sum_{j = c, \bar{c}} K^{-1}_{ij} \otimes\bigg[
  f_j^{\nfour}(Q^2)  - \sum_{k,l = g, q, \bar{q}} K_{jk}\left( \smallfrac{m_c^2}{Q^2}\right)\otimes \tilde{K}^{-1}_{kl}\left( \smallfrac{m_c^2}{Q^2}\right)\otimes f_l^{\nfour}(Q^2)\bigg],
\label{eq:ICPDF}
\eea
where in the second line we used Eq.~\eqref{eq:Kinvcq} in 
Appendix \ref{sec-appendixmat}, or equivalently solved
Eqs.~(\ref{eq:matchpdfq}, \ref{eq:matchpdfc}) for $f_i^{\nthree}$. We
thus find that
\bea
\Delta F^{\nthree} (Q^2,m_c^2)
&=\sum_{i = c, \bar{c}}\bigg[
  C_i^{\nthree} \left( \smallfrac{m_c^2}{Q^2}\right)
 - \sum_{k = g, q, \bar{q}} 
B^{\nfour}_k \left( \smallfrac{m_c^2}{Q^2}\right)
\otimes K_{ki}\left( \smallfrac{m_c^2}{Q^2}\right)\bigg]
\nonumber\\
&\qquad\otimes
\sum_{j = g, q,\bar q, c, \bar{c}}K^{-1}_{ij}\left( \smallfrac{m_c^2}{Q^2}\right)\otimes f_j^{\nfour}(Q^2)
 .\label{eq:FthreeIC4}
\eea
Using Eq.~\eqref{eq:FthreeIC4} in the generic FONLL formula Eq.~\eqref{eq:FONLL} then gives 
\begin{equation}
F_{\rm FONLL} (Q^2,m_c^2)=F_{\rm FONLL}(Q^2,m_c^2)\Big|_{\rm zic}+\Delta F_{\rm FONLL} (Q^2,m_c^2) ,
\label{eq:FONLLICsep}
\end{equation}
where the intrinsic charm contribution 
\begin{align}
&\Delta F_{\rm FONLL} \left(Q^2,m_c^2 \right) =  
 \sum_{i,j = c, \bar{c}}\bigg[
  \Big(C_i^{\nthree}\left( \smallfrac{m_c^2}{Q^2}\right)-C_i^{\nthreez}\left( \smallfrac{m_c^2}{Q^2}\right)\Big)
\nonumber\\ &\qquad\qquad\qquad\qquad\qquad\qquad
- \sum_{m = g, q, \bar{q}} 
\Big(B^{\nfour}_m\left( \smallfrac{m_c^2}{Q^2}\right)- B^{\nfourz}_m\left( \smallfrac{m_c^2}{Q^2}\right)\Big)\otimes K_{mi}\left( \smallfrac{m_c^2}{Q^2}\right)\bigg]\nonumber\\
&\qquad
\otimes K^{-1}_{ij}\left( \smallfrac{m_c^2}{Q^2}\right)\otimes \bigg[
  f_j^{\nfour}(Q^2)  - \sum_{k,l = g, q, \bar{q}} K_{jk}\left( \smallfrac{m_c^2}{Q^2}\right)\otimes \tilde{K}^{-1}_{kl}\left( \smallfrac{m_c^2}{Q^2}\right)\otimes f_l^{\nfour}(Q^2)\bigg]\, .
\label{eq:FONLLIC}
\end{align} 
At large $Q^2$ this term is manifestly $\Ord(m_c^2/Q^2)$, so has no
effect on the high energy limit. Near threshold it is in general
$\Ord(1)$, unless the intrinsic charm vanishes: then using
Eq.~\eqref{eq:fthreecord} it is easy to see that
$\Delta F_{\rm FONLL}$ is subleading (down by one power of
$\alpha_s(Q^2)$).

We stress that Eq.~\eqref{eq:FthreeICsep}, with the two terms given by
Eqs.~(\ref{eq:FONLLzIC}, \ref{eq:FONLLIC}), is actually identical to
the simple expression Eq.~\eqref{eq:ACOT2} order by order in
perturbation theory. In a sense this is obvious because both equations
have been derived with the FONLL construction, Eq.~\eqref{eq:FONLL},
systematically rewriting expressions involving 3FS PDFs in
terms of 4FS PDFs using the matching conditions
Eq.~\eqref{eq:matchpdf}. This said, it is perhaps useful, if only as a
cross check, to verify by explicit computation that the two expression
are indeed identical.

To do this, it is clearly sufficient to show that
Eq.~\eqref{eq:FthreeICsep}, with the two terms given by
Eq.~\eqref{eq:FthreezICB} and Eq.~\eqref{eq:FthreeIC4}, is equivalent
to Eq.~\eqref{eq:Fthreefour}. We first rewrite it, by collecting the
coefficients of $f^{\nfour}_i$ (and suppressing the arguments of the
functions to lighten the notation), as
\begin{align}
  F^{\nthree}
&= \sum_{k = g, q, \bar{q}} \Big[ B^{\nfour}_k + \sum_{i = c,\bar c} \Big(C_i^{\nthree} - \sum_{j = g, q, \bar{q}} B^{\nfour}_j \otimes K_{ji} \Big) \otimes K^{-1}_{ik}
 \Big]\otimes f_k^{\nfour}\nonumber\\
&\qquad + \sum_{i,j = c,\bar c} \Big(C_i^{\nthree} - \sum_{k = g, q, \bar{q}} B^{\nfour}_k \otimes K_{ki} \Big) \otimes K^{-1}_{ij} \otimes f^{\nfour}_j\nonumber\\
&= \sum_{k = g, q, \bar{q}} \bigg[ \sum_{i = g, q, \bar{q}} C_i^{\nthree} \otimes 
\Big(\tilde{K}_{ik}^{-1} - \sum_{j = g, q, \bar{q}}\sum_{l = c, \bar{c}} \tilde{K}_{ij}^{-1}\otimes K_{jl} \otimes K^{-1}_{lk}\Big)
+\sum_{i = c, \bar c} C_i^{\nthree}  \otimes K_{ik}^{-1}\bigg] \otimes f_k^{\nfour}\nonumber\\
&\qquad + \sum_{k = c, \bar c} \bigg[\sum_{i = c, \bar c} C_i^{\nthree}  \otimes K_{ik}^{-1} - 
\sum_{i,j = g, q, \bar{q}}\sum_{l = c, \bar c} C_i^{\nthree} \otimes \tilde{K}_{ij}^{-1}\otimes K_{jl}\otimes K^{-1}_{lk} \bigg] \otimes f^{\nfour}_k,
\end{align}
using Eq.~\eqref{eq:Bdef}. Now using the expressions in
App.~\ref{sec-appendixmat}, specifically
Eqs.~(\ref{eq:Kinvcq}, \ref{eq:Kinvqq}) in the first line and
Eq.~\eqref{eq:Kinvqc} in the second, we find
\begin{align}
F^{\nthree} &= \sum_{k = g, q, \bar{q}} \bigg[ \sum_{i = g, q, \bar{q}} C_i^{\nthree} \otimes K_{ik}^{-1} 
+\sum_{i = c, \bar c} C_i^{\nthree}  \otimes K_{ik}^{-1}\bigg] \otimes f_k^{\nfour}\nonumber\\
&\quad + \sum_{k = c,\bar c} \bigg[\sum_{i = g, q, \bar{q}} C_i^{\nthree} \otimes K_{ik}^{-1} 
+\sum_{i = c, \bar c} C_i^{\nthree}  \otimes K_{ik}^{-1} \bigg] \otimes f^{\nfour}_k \nonumber\\
&=  \sum_{i,k = g, q, \bar{q},c,\bar{c}} C^{\nthree}_i \otimes K^{-1}_{ik}\otimes f^{\nfour}_k,
\end{align}
which is the desired result, Eq.~\eqref{eq:Fthreefour}.

It follows that the formulation of FONLL in Ref.~\cite{Forte:2010ta},
Eq.~\eqref{eq:FONLLzIC}, in which it is assumed that all charm is
generated perturbatively, plus the extra `intrinsic charm'
contribution Eq.~\eqref{eq:FONLLIC}, is identical to the full ACOT
result through Eq.~\eqref{eq:ACOT2}, irrespective of any condition on
the charm at the initial scale.

Note that this means that we can write the full FONLL (or equivalently
ACOT) expression Eq.~(\ref{eq:FONLLICsep}) as simply
\begin{align}
F_{\rm FONLL} \left(Q^2,m_c^2 \right) &= \sum_{i = g, q, \bar{q}}
  B^{\nfour}_i \otimes f_i^{\nfour} (Q^2) \nonumber\\
&+\sum_{i = c, \bar{c}}\Big[
  C_i^{\nthree} - \sum_{k = g, q, \bar{q}} B^{\nfour}_k\otimes K_{ki}\Big]
\otimes \sum_{j = g,q,\bar q, c, \bar{c}} K^{-1}_{ij} \otimes f_j^{\nfour} ,
\label{eq:FONLLsimp}
\end{align} 
which follows directly from Eq.~\eqref{eq:F3derivation} and the
observation made in Sect.~\ref{sec:ACOT}
that all the massless contributions (and thus the difference term in Eq.~\eqref{eq:FONLLd})
cancel when no assumption is made about instrinsic charm. The second
term is now essential to obtain the correct high $Q^2$ behaviour, even
in the limit of zero intrinsic charm.
The diagrams contributing to the full FONLL result Eq.~\eqref{eq:FONLLsimp}
are shown schematically in Fig.~\ref{fig:counting} (lower row).

Comparison of the ACOT representation Eq.~(\ref{eq:ACOT}) with the
original zero intrinsic charm representation of FONLL
Eq.~(\ref{eq:FONLLzIC}) gives us a new way to understand the origin of
intrinsic charm contribution Eq.~(\ref{eq:FONLLIC}): taking the
difference, and using the definitions Eq.~(\ref{eq:Bdef}) and
Eq.~(\ref{eq:Bdef0})
\begin{align}
\Delta F_{\rm FONLL} 
&= \sum_{i,j = g, q, \bar{q}, c,\bar{c}}\left[C^{\nthree}_i-
C_i^{\nthreez}\right]\otimes K_{ij}^{-1}\otimes
  f_j^{\nfour} - \sum_{i,j = g, q, \bar{q}}\left[C^{\nthree}_i-C_i^{\nthreez}\right]\otimes \tilde{K}_{ij}^{-1}\otimes
  f_j^{\nfour}  \nonumber\\
&= \sum_{i,j = g, q, \bar{q}, c,\bar{c}}\left[C^{\nthree}_i-
C_i^{\nthreez}\right]\otimes \left[K_{ij}^{-1}- \widehat{K_{ij}^{-1}}\right]\otimes
  f_j^{\nfour}   ,
\label{eq:DeltaFproj}
\end{align}
where the matrix $\widehat{K_{ij}^{-1}}$ acts as a projector onto the space of light partons
\begin{equation}
\widehat{K_{ij}^{-1}} = \begin{cases}\tilde{K}_{ij}^{-1}&\qquad i,j= g, q, \bar{q}\\
                            0, &\qquad {\rm otherwise.} \end{cases}\label{eq:Kproj}
\end{equation}
The expression Eq.~(\ref{eq:DeltaFproj}) is particularly transparent:
when intrinsic charm is included the massless coefficient functions in
the charm sector must be mass corrected, with additional collinear subtractions for
the incoming charm quarks lines, these subtractions being factorized
multiplicatively.

From Eq.~\eqref{eq:DeltaFproj} it is also possible to derive another
useful form of $\Delta F_{\rm FONLL}$ in terms of 4FS coefficient functions.
Substituting the inverse of Eq.~(\ref{eq:C4C3K}),
\begin{equation}
  C_i^{\nthree} \left(\smallfrac{m_c^2}{Q^2}\right)
=\sum_{j = g, q, \bar{q}, c, \bar{c}} C_j^{\nfour}\left(\smallfrac{m_c^2}{Q^2}\right)\otimes
  K_{ji}\left( \smallfrac{m_c^2}{Q^2}\right) ,\label{eq:C3C4K} 
\end{equation}
and Eq.~\eqref{eq:masslimdef} into Eq.~\eqref{eq:DeltaFproj} we find immediately
\begin{equation}
\Delta F_{\rm FONLL} \left(Q^2,m_c^2 \right)=  
 \sum_{i = g, q, \bar{q}, c, \bar{c}}
  \Big[C_i^{\nfour}\left( \smallfrac{m_c^2}{Q^2}\right)-C_i^{\nfour}\left( 0\right)\Big]\otimes
\Big[f_i^{\nfour}  - \sum_{k,l = g, q, \bar{q}} 
K_{ik}\otimes \tilde{K}^{-1}_{kl}\otimes f_l^{\nfour}\Big].
\label{eq:FONLLIC4tmp}
\end{equation}
When $i=g, q, \bar{q}$ in the sum, the difference in the second square brackets vanishes,
because of Eq.~\eqref{eq:Ktilinvgg}.
Therefore Eq.~\eqref{eq:FONLLIC4tmp} simplifies to
\begin{equation}
\Delta F_{\rm FONLL} \left(Q^2,m_c^2 \right)=  
 \sum_{i = c, \bar{c}}
  \Big[C_i^{\nfour}\left( \smallfrac{m_c^2}{Q^2}\right)-C_i^{\nfour}\left( 0\right)\Big]\otimes
\Big[f_i^{\nfour}  - \sum_{k,l = g, q, \bar{q}} 
K_{ik}\otimes \tilde{K}^{-1}_{kl}\otimes f_l^{\nfour}\Big].
\label{eq:FONLLIC4}
\end{equation} 
This is a very compact expression, and manifestly shows that the missing
mass corrections in Eq.~\eqref{eq:FONLLzIC} due to intrinsic charm are entirely
determined by the mass dependence of the charm initiated contribution in the 4FS.

\subsection{Comparison to S-ACOT}
\label{sec:SACOT}

Finally, we consider the connection to S-ACOT \cite{Collins:1997sr,Kramer:2000hn,Guzzi:2011ew},
a simplified variant of ACOT whose validity is based on the assumption that the charm is generated perturbatively.
Under this assumption, the authors of Ref.~\cite{Kramer:2000hn} claim that
in the construction of the ACOT (massive 4FS) coefficient functions
the mass dependence in all diagrams with an
incoming charm quark can be systematically ignored, i.e.\
$C_{c,\bar{c}}^{\nfour}(m_c^2/Q^2)$ can be replaced with
$C_{c,\bar{c}}^{\nfour}(0)$ in all steps of the construction.

More precisely, the structure functions in S-ACOT are written as in Eq.~\eqref{eq:Ffour},
\beq\label{eq:S-ACOT}
F_{\text{S-ACOT}}(Q^2,m_c^2) = \sum_{i=g,q,\bar q,c,\bar c}
\bar C^{\nfour}_i \(\smallfrac{m_c^2}{Q^2}\) \otimes f^{\nfour}_i(Q^2)
\eeq
but with new coefficient functions $\bar C^{\nfour}_i$. Those must be determined by consistency with
the unresummed result, Eq.~\eqref{eq:Fthree}; using Eq.~\eqref{eq:matchpdf} we find 
\beq\label{eq:Cbar}
C^{\nthree}_i\(\smallfrac{m_c^2}{Q^2}\) = \sum_{i=g,q,\bar q,c,\bar c}
\bar C^{\nfour}_j \(\smallfrac{m_c^2}{Q^2}\) \otimes K_{ji} \(\smallfrac{m_c^2}{Q^2}\).
\eeq
In the general case which can account for intrinsic charm,
this has a unique solution, $\bar C^{\nfour}_i=C^{\nfour}_i$ for $i=g,q,\bar{q},c,\bar{c}$, giving back ACOT.
However, in the absence of intrinsic charm, $f^{\nthree}_{c,\bar c}=0$, and thus 
Eq.~\eqref{eq:Cbar} can only be derived for $i=g,q,\bar{q}$.
This means that the system of Eq.~\eqref{eq:Cbar} is under-constrained,
in the sense that solving it for $\bar C^{\nfour}_j (m_c^2/Q^2)$ is ambiguous:
there are only $7$ equations for $9$ unknowns. This is another manifestation of the ambiguity
in inverting Eq.~\eqref{eq:fcfourthreezic} discussed in Sect.~\ref{sec:ECIC}.
When this is the case, we are free to choose two of the coefficient functions 
$\bar C^{\nfour}_j (m_c^2/Q^2)$ as we please, subject only to the constraint that
we recover the massless coefficient functions when $Q^2\gg m_c^2$.
The most natural choice is then the S-ACOT simplification~\cite{Kramer:2000hn}
\beq\label{eq:CbarSACOTc}
\bar C^{\nfour}_{i}\(\smallfrac{m_c^2}{Q^2}\) = C^{\nfour}_{i}(0), \qquad i=c,\bar c.
\eeq
Given this, we can then solve Eq.~\eqref{eq:Cbar} for the remaining components, giving immediately
\beq\label{eq:CbarSACOT}
\bar C^{\nfour}_{i}\(\smallfrac{m_c^2}{Q^2}\) = \sum_{j=g,q,\bar q}
\bigg[C^{\nthree}_{i}\(\smallfrac{m_c^2}{Q^2}\) - \sum_{k=c,\bar c} C^{\nfour}_{k}(0) \otimes K_{kj}
\bigg] \otimes \tilde K^{-1}_{ji}
, \qquad i=g,q,\bar q.
\eeq
These are the S-ACOT coefficient functions, which can be determined order by order in perturbation
theory starting from the massive 3FS coefficients for light channels and the massless
4FS coefficients for the heavy channel.
The physical interpretation is very transparent: the collinear logarithms due to the charm quark
are completely subtracted off $C^{\nthree}_{i}$, while power suppressed contributions are all left untouched.

We now want to compare this result to FONLL.
We saw in Sect.~\ref{sec:fonll-zic} that when all charm is generated
perturbatively, Eq.~\eqref{eq:FONLLzIC} does not depend on
$C_{c,\bar{c}}^{\nfour}(m_c^2/Q^2)$: in the
full FONLL expression Eq.~(\ref{eq:FONLLICsep}) (which we just showed
is equivalent to ACOT) all the mass dependence of the incoming charm
quark lines is contained in the 4FS coefficient functions
$C_{c,\bar{c}}^{\nfour}(m_c^2/Q^2)$ in the
$\Delta F_{\rm FONLL}$ term Eq.~(\ref{eq:FONLLIC4}).
When we set
$C_{c,\bar{c}}^{\nfour}(m_c^2/Q^2)\to
C_{c,\bar{c}}^{\nfour}(0)$ as in S-ACOT,
$\Delta F_{\rm FONLL}$ vanishes identically,
whether or not we have intrinsic charm (i.e.\ whether or not the
PDF term in square brackets in Eq.~\eqref{eq:FONLLIC4} vanishes). 
It follows that S-ACOT is equivalent order by order in
perturbation theory to FONLL as formulated in Ref.~\cite{Forte:2010ta}
with all charm generated perturbatively,
Eq.~(\ref{eq:FONLLzIC}). This of course accounts for the
numerical equivalence of FONLL and S-ACOT at NLO discovered in
Ref.~\cite{Binoth:2010nha}.\footnote
{There are other sources of differences at finite order between FONLL and S-ACOT,
due to a damping factor adopted in Ref.~\cite{Forte:2010ta} and the $\chi$ rescaling
sometimes used in practical applications of S-ACOT. See discussion in Sect.~\ref{sec:damping}.}

The equivalence between FONLL with zero intrinsic charm Eq.~\eqref{eq:FONLLzIC}
and S-ACOT Eq.~\eqref{eq:S-ACOT} implies the relation
\beq\label{eq:CbarFONLL}
\bar C^{\nfour}_{i}\(\smallfrac{m_c^2}{Q^2}\) = 
B^{\nfour}_i \left( \smallfrac{m_c^2}{Q^2}\right)-B^{\nfourz}_i \left( \smallfrac{m_c^2}{Q^2}\right)+C_i^{\nfour}(0),
\eeq
for $i=g,q,\bar{q}$. It is straightforward to check that Eq.~\eqref{eq:CbarFONLL} is identical to Eq.~\eqref{eq:CbarSACOT},
as expected.
It is interesting to observe that while in S-ACOT the subtraction of charm-induced collinear logarithms
is achieved identifying the logarithms in a factorized form, in the FONLL formulation
the collinear logarithms are subtracted using the `massless limit' $B^{\nfourz}_i$ which includes
also constant (non-log) terms, which are then restored through $C_i^{\nfour}(0)$.

In summary, we have shown that
\begin{align}
F_{\text{ACOT}}  &\equiv  F_{\text{FONLL}}\label{eq:ACOTeqFONLL}\\
F_{\text{S-ACOT}}  &\equiv  F_{\text{FONLL}}\Big|_{\rm zic}\label{eq:SACOTeqZIC}
\end{align}
to all orders in perturbation theory. The first of these equivalences
is a direct consequence of the fact that both ACOT and FONLL express
their final result in terms of PDFs factorized in the 4FS:
since the PDFs are then formally identical, the coefficient
functions must also be identical, order by order in perturbation
theory. The second equivalence is more subtle: it states that
suppressing the mass dependence in the coefficient functions with
incoming charm is actually equivalent to the simplified result obtained when there is no intrinsic charm
PDF, where suppressed contributions proportional to Eq.~\eqref{eq:fthreecord} are neglected.
It is particularly useful, since it means that it is unnecessary
to compute massive coefficient functions with incoming charm if all
charm is generated perturbatively: in this situation S-ACOT is
exact. However reliable calculations with a fitted charm distribution
do require knowledge of these coefficient functions: in this
circumstance S-ACOT can only be an approximation (and not necessarily
a very reliable one \cite{Ball:2015tna}).

\subsection{Damping factor}
\label{sec:damping}

Our discussion so far was mostly formal, focussing on all-order expressions.
When the various contributions are computed at finite order, higher order interference terms may spoil
the accuracy of the results, as discussed in Sect.~\ref{sec:matching}.
To avoid this problem, the computations in Ref.~\cite{Forte:2010ta} also include a
phenomenological damping factor: in place of Eq.~(\ref{eq:FONLLd}) one writes
\begin{align}
F_{\rm FONLL} (Q^2,m_c^2)
&= F^{\nthree}( Q^2,m_c^2)+ D\left(\smallfrac{m_c^2}{Q^2}\right)\left[ F^{\nfour} (Q^2,0)
-F^{\nthreez}(Q^2,m_c^2)\right]\nonumber\\
&\equiv F^{\nthree}( Q^2,m_c^2)+ D\left(\smallfrac{m_c^2}{Q^2}\right) F^{(d)} (Q^2,m_c^2 )\, ,
\label{eq:FONLLdamp}
\end{align} 
where
\begin{equation}
D\left(\smallfrac{m_c^2}{Q^2}\right) = \Theta (Q^2-m_c^2)\(1-\smallfrac{m_c^2}{Q^2}\)^2
\label{eq:damping}
\end{equation}
suppresses the difference term $F^{(d)}$ close to threshold, i.e.\ when $Q^2\sim m_c^2$.
This is the region where the resummation of collinear logarithms, added to the fixed-order
result $F^{\nthree}( Q^2,m_c^2)$ through the difference term $F^{(d)} (Q^2,m_c^2 )$,
is not needed and can therefore be artificially suppressed.
This suppression turns out to be important when working at $\Ord(\as)$,
where the $\Ord(\as^2)$ interference terms are sizeable, but becomes almost negligible
already at $\Ord(\as^2)$, where the $\Ord(\as^2)$ interference terms are small.

With this damping Eq.~(\ref{eq:FONLLzIC}) becomes
\begin{align}
& F_{\rm FONLL} (Q^2,m_c^2)\Big|_{\rm zic}
  =  \sum_{i = g, q, \bar{q}}
  B^{\nfour}_i \left( \smallfrac{m_c^2}{Q^2}\right)\otimes f_i^{\nfour} (Q^2)\nonumber\\
  &\qquad + D\left(\smallfrac{m_c^2}{Q^2}\right)
    \bigg[\sum_{i = g, q, \bar{q}}
  \(C_i^{\nfour}(0)- B^{\nfourz}_i\left( \smallfrac{m_c^2}{Q^2}\right)\)\otimes f_i^{\nfour} (Q^2) + \sum_{i = c, \bar{c}}
  C_i^{\nfour}(0)\otimes f_i^{\nfour} (Q^2)\bigg]
 \label{eq:FONLLzICdamp}
\end{align}
and, likewise, Eq.~(\ref{eq:FONLLIC}) becomes
\begin{align}
\Delta F_{\rm FONLL} (Q^2,m_c^2) &=  
 \sum_{i= c, \bar{c}}\bigg[
  \(C_i^{\nthree}\left( \smallfrac{m_c^2}{Q^2}\right)-D\left(\smallfrac{m_c^2}{Q^2}\right) C_i^{\nthreez}\left( \smallfrac{m_c^2}{Q^2}\right)\)
\nonumber\\ &\qquad\qquad- \sum_{m = g, q, \bar{q}} 
\(B^{\nfour}_m\left( \smallfrac{m_c^2}{Q^2}\right)- D\left(\smallfrac{m_c^2}{Q^2}\right)B^{\nfourz}_m\left( \smallfrac{m_c^2}{Q^2}\right)\)
              \otimes K_{mi}\left( \smallfrac{m_c^2}{Q^2}\right)\bigg]
\nonumber\\
&\qquad \otimes \sum_{j=g,q,\bar q,c,\bar c} {K}^{-1}_{ij}\left( \smallfrac{m_c^2}{Q^2}\right)\otimes f_j^{\nfour}(Q^2)\, .
\label{eq:FONLLICdamp}
\end{align}
However since when when add $\Delta F_{\rm FONLL}$ to
$F_{\rm FONLL}\big|_{\rm zic}$ the difference term vanishes identically
(and thus for example in Eq.~(\ref{eq:FONLLsimp}) there are no massless
terms to damp), it is clear that when there is intrinsic charm the damping has no effect whatsoever.

It may seem paradoxical that while the limit of zero
intrinsic charm should be unique, the zero intrinsic charm result
Eq.~(\ref{eq:FONLLzICdamp}) clearly depends on the arbitrary function $D$. The reason of
course is that in taking the limit the $\Delta F$ term
Eq.~(\ref{eq:FONLLICdamp}) is suppressed, since it becomes formally
subleading: although small, it is still not entirely negligible, and indeed
must be similar in size to the subleading variation achieved through
changing the damping factor $D$.  The FONLL damping factor is thus another
manifestation of the ambiguity in the treatment of the zero
intrinsic charm limit, discussed in the previous Section. 

The $\chi$-rescaling prescription plays a similar role in S-ACOT-$\chi$.
In this case, however, rather than damping the (massless) resummation contribution,
the massive kinematics is restored in those contributions which are computed in the massless limit.
In this way, the S-ACOT-$\chi$ result becomes closer to ACOT, since (the dominant) part
of the neglected power corrections are reinstated.
This means that, even in the presence of intrinsic charm, S-ACOT-$\chi$ can be a reasonable approximation.
Given the restriction of the exact massive results for the charm initiated contributions to $\Ord(\as)$ (the $\Ord(\as^2)$ diagrams with incoming massive quark lines have yet to be calculated),
the usage of S-ACOT-$\chi$ might be a useful tool for improving the accuracy of calculations in the presence of intrinsic charm to $\Ord(\as^2)$ and beyond
(see e.g.\ Ref.~\cite{Stavreva:2012bs}).

%% file: sections/sec-beauty.tex
\section{From Charm to Bottom and Top}
\label{sec:bt}

So far in this paper we have not considered the third generation quarks,
ignoring in particular top and bottom mass dependence in the coefficient functions.
This means in effect that we assumed that top and bottom were infinitely heavy,
so that we were always well below threshold for their production.
Virtual effects are suppressed by powers of the quark mass, provided a 
decoupling renormalization scheme is used for these quarks.

In practice, the bottom quark is not much heavier than the charm quark,
so, while the top quark can be safely ignored at small scales,
considering the bottom quark to be infinitely heavy is not a very good approximation even at the charm threshold. Virtual bottom quark loops cost a power of $\alpha_s$ in a gluon propagator, so bottom mass effects appear first at NNLO.
Their effects are thus small, but not completely negligible.

It is the purpose of this Section to extend the discussion in Sect.~\ref{sec:fact} and Sect.~\ref{sec:fonll}
to include the bottom and top quarks, in a complete and coherent framework.
We first concentrate on the bottom quark, and then generalise to the top quark.
We will also discuss the possibility of intrinsic beauty.

\subsection{The bottom quark}
\label{sec:beauty}

In both the schemes (3FS and 4FS) discussed in Sect.~\ref{sec:fact}, 
bottom quark effects can appear in the perturbative coefficient functions through additional diagrams, either as virtual loops or, when kinematically allowed, through pair production.
In both schemes the UV divergences due to bottom loops are renormalized in the decoupling (CWZ)
scheme, so the bottom mass effects are formally suppressed as $Q^2/m_b^2$ when $Q^2\ll m_b^2$. However in practice this condition is never really satisfied, and while pair production vanishes below threshold, the effect of virtual bottom quark loops should be included at NNLO and beyond. 

Hence, all coefficient functions discussed so far implicitly include a dependence on
the bottom mass. We thus replace
\bea
C_i^{\nthree}\(\smallfrac{m_c^2}{Q^2}\)& \to  C_i^{\nthree} \(\smallfrac{m_c^2}{Q^2},\smallfrac{m_b^2}{Q^2}\) \\
C_i^{\nfour}\(\smallfrac{m_c^2}{Q^2}\)& \to  C_i^{\nfour} \(\smallfrac{m_c^2}{Q^2},\smallfrac{m_b^2}{Q^2}\)
\label{eq:replacement_b}
\eea
and equivalently the massless limit $C_i^{\nfour}(0) \to  C_i^{\nfour} (0,m_b^2/Q^2)$;
the same extension applies also for derived quantities such as $B^{\nfour}_i$ or $C_i^{\nthreez}$. The $m_b$ dependence is computed at fixed order: below threshold it can appear at $\Ord(\as^2)$ through a 1-loop correction to a gluon propagator, while above threshold it will appear already as an $\Ord(\as)$ contribution to the structure function (e.g.\ $F_2$) through the production of bottom quarks in the final state (see Fig.~\ref{fig:counting}). 

Since one naturally performs the resummation of charm collinear logarithms
first, the 4FS plays for beauty the same role that the 3FS plays for charm.
Entirely analogous to Eq.~(\ref{eq:Fthree}) we thus have
\begin{equation}
  F^{\nfour} (Q^2,m_c^2,m_b^2) =  \sum_{i = g,q', \bar{q}',b ,\bar{b}}
  C_i^{\nfour} \left( \smallfrac{m_c^2}{Q^2},\smallfrac{m_b^2}{Q^2} \right)\otimes
  f_i^{\nfour} (Q^2) \label{eq:Ffourb}
\end{equation}
where now $q'=d,u,s,c$, and we have included the theoretical possibility of  intrinsic beauty through the addition of a $b$-quark PDF.
In this expression all the large logarithms $\log (m_c^2/Q^2)$ 
have been resummed, but at large $Q^2$ the potentially large logarithms $\log (m_b^2/Q^2)$
remain unresummed in the coefficient functions $C_i^{\nfour}(m_c^2/Q^2,m_b^2/Q^2)$. 
The 4FS PDF evolution Eq.~\eqref{eq:Efour} can be trivially extended to include the bottom quark PDF as
\begin{equation}
f^{\nfour}_i(x,Q^2)= \sum_{j = g,q', \bar{q}',b,\bar{b}}
\bigGamma^{\nfour}_{ij}\left(Q^2,Q_0^2\right)\otimes f^{\nfour}_j(Q_0^2);\label{eq:Efourb}
\end{equation}
where 
\begin{equation}
\bigGamma_{ij}^{\nfour}(Q^2,Q_0^2) = \begin{cases}\Gamma_{ij}^{\nfour}(Q^2,Q_0^2)&\qquad i,j=g,q', \bar{q}' \\
\delta_{ij}, &\qquad {i,j=b, \bar{b}}\\ 
0, &\qquad {\rm otherwise.} \end{cases}\label{eq:Gamfour}
\end{equation}
The $Q^2$ dependence of the $b$ contribution to the structure function is all in the coefficient function, so $f_{b ,\bar{b}}^{\nfour}$ is independent of $Q^2$.
Note that it is precisely the fact that the $b$ quarks do not mix with the
lighter partons in the 4FS that makes the extension of the previous formalism to
include bottom quark effects trivial. If we wish to assume that there is no intrinsic beauty, we can simply take
\begin{equation} 
f_b^{\nfour}= f_{\bar{b}}^{\nfour} = 0, \label{eq:zibfour}
\end{equation}
in analogy with Eq.~\eqref{eq:zicthree} for no intrinsic charm. 

At high scales $Q^2\gg m_b^2$, the logarithms of $m_b^2/Q^2$ in $C_i^{\nfour}$
become large and need to be resummed.
We must therefore factorize the large logarithms due to the bottom quark into the PDF, just as we did for the charm,
leading naturally to a 5 flavor scheme (5FS). PDFs in the 5FS evolve as
\begin{equation}
f^{\nfive}_i(Q^2)=\sum_{j = g, q', \bar{q}', b ,\bar{b}}\Gamma^{\nfive}_{ij}\left(Q^2,Q_0^2\right)\otimes f^{\nfive}_j(Q_0^2).\label{eq:Efive}
\end{equation}
and are related to 4FS PDFs by matching conditions analogous to Eq.~\eqref{eq:matchpdf},
\beq
f_i^{\nfive} (\mub^2) 
= \sum_{j = g, q', \bar{q}', b, \bar{b}} K_{ij}^{\nff}\( \smallfrac{m_b^2}{\mub^2}\)\otimes f_j^{\nfour} (\mub^2),
\label{eq:matchpdf5}
\eeq
where we have introduced new matching functions $K_{ij}^{\nff}$, using a label ${}^{\nff}$
to distinguish them from the previous $K_{ij}\equiv K_{ij}^{\nft}$:
in practice they are the same quantities, except that there is one more active flavour, and the `heavy' index is now $b$.
The scale $\mub\sim m_b$ is the threshold scale at which the 4FS PDFs are converted into 5FS PDF,
during perturbative evolution.

In analogy with the ACOT expression Eq.~\eqref{eq:ACOT}, equivalent to FONLL when there is intrinsic beauty, the resummed result can be thus be written as
\bea
F_{\rm ACOT}(Q^2,m_c^2,m_b^2) &= \sum_{i,j = g, q', \bar{q}', b, \bar{b}}
C_i^{\nfour} \left(\smallfrac{m_c^2}{Q^2},\smallfrac{m_b^2}{Q^2} \right)\otimes
{K_{ij}^{\nff}}^{-1} \left( \smallfrac{m_b^2}{Q^2}\right)\otimes
f_j^{\nfive} (Q^2) \nonumber\\
 &= \sum_{i,j = g, q', \bar{q}', b, \bar{b}}
C_i^{\nfive} \left(\smallfrac{m_c^2}{Q^2},\smallfrac{m_b^2}{Q^2} \right)\otimes
f_j^{\nfive} (Q^2),
\eea
or equivalently, using the FONLL construction Eqs.~(\ref{eq:FONLLzIC}, \ref{eq:FONLLIC}), as
\begin{align}
&F_{\rm FONLL}(Q^2,m_c^2,m_b^2)\nonumber\\
  &\quad=  \sum_{i = g, q', \bar{q}'}
    \[B^{\nfive}_i \left( \smallfrac{m_c^2}{Q^2},\smallfrac{m_b^2}{Q^2}\right)
    -B^{\nfivez}_i \left( 0,\smallfrac{m_b^2}{Q^2}\right)
    +C_i^{\nfive}\(0,0\)\]
    \otimes f_i^{\nfive} (Q^2)\nonumber\\
  &\qquad + \sum_{i = b, \bar{b}}
    C_i^{\nfive}\(0,0\)\otimes f_i^{\nfive} (Q^2)
    \nonumber\\
  &\qquad + \sum_{i,j = b, \bar{b}}\bigg[
    C_i^{\nfour} \( \smallfrac{m_c^2}{Q^2},\smallfrac{m_b^2}{Q^2}\)
    -C_i^{\nfourz}\(0,\smallfrac{m_b^2}{Q^2}\)
    \nonumber\\
  &\qquad\qquad\qquad
    - \sum_{m = g,q', \bar{q}'} 
    \bigg(B^{\nfive}_m\left( \smallfrac{m_c^2}{Q^2},\smallfrac{m_b^2}{Q^2}\right)
    - B^{\nfivez}_m\left(0,\smallfrac{m_b^2}{Q^2}\right)\bigg)
    \otimes K_{mi}^{\nff}\left( \smallfrac{m_b^2}{Q^2}\right)\bigg]
    \nonumber\\
  &\qquad\quad
    \otimes {K^{\nff}_{ij}}^{-1}\(\smallfrac{m_b^2}{Q^2}\)\otimes\Big[
  f_j^{\nfive}(Q^2)  - \sum_{k,l = g,q', \bar{q}'} 
K_{jk}^{\nff}\(\smallfrac{m_b^2}{Q^2}\)\otimes \tilde K{{}^{\nff}_{kl}}^{-1}\(\smallfrac{m_b^2}{Q^2}\)\otimes f_l^{\nfive}(Q^2)\Big]\, ,
\label{eq:FONLLB}
\end{align}
where $\tilde{K}{{}^{\nff}_{ji}}^{-1}$ is the inverse of
$K_{ij}^{\nff}$ restricted to the subspace of $i,j=g, q', \bar{q}'$.
Note that although the massive coefficient functions now depend on
both $m_c^2$ and $m_b^2$, all the matching matrices depend only
on $m_b^2/Q^2$, since all the large logarithms of $m_c^2/Q^2$ were
resummed in the previous step. Furthermore in the massless terms, we set both $m_b$ and $m_c$ to zero, since $m_b^2/Q^2 > m_c^2/Q^2$: the corresponding definitions of the massless subtractions are thus
\bea
B^{\nfive}_i\left(\smallfrac{m_c^2}{Q^2}, \smallfrac{m_b^2}{Q^2} \right) &= 
\sum_{j = g, q', \bar{q}'}
  C_j^{\nfour} \left(\smallfrac{m_c^2}{Q^2}, \smallfrac{m_b^2}{Q^2}\right)\otimes
  \tilde{K}{{}^{\nff}_{ji}}^{-1} \(\smallfrac{m_b^2}{Q^2}\)
\label{eq:Bfourb}\\
B^{\nfivez}_i\left(0, \smallfrac{m_b^2}{Q^2} \right) &= 
\sum_{j = g, q', \bar{q}'}
  C_j^{\nfourz} \left(0, \smallfrac{m_b^2}{Q^2}\right)\otimes
  \tilde{K}{{}^{\nff}_{ji}}^{-1} \(\smallfrac{m_b^2}{Q^2}\)
\label{eq:Bfourzb}
\eea
where $C_j^{\nfourz}(0,m_b^2/Q^2)$ is the singular massless limit of $C_j^{\nfour}(0,m_b^2/Q^2)$, which can be written as
\beq
C_j^{\nfourz}\left(0, \smallfrac{m_b^2}{Q^2} \right)
=\sum_{j = g, q', \bar{q}', b, \bar b} C_j^{\nfive}\left(0,0 \right)\otimes
  K_{ji}^{\nff}\left( \smallfrac{m_b^2}{Q^2}\right) .
\label{eq:C40}
\eeq
This last equation then gives the correct definition of the
matching coefficients $K_{ij}^{\nff}( m_b^2/Q^2)$, corresponding to
Eq.~\eqref{eq:masslimdef}. Note that the subtraction of
$B^{\nfivez}_i(0,m_b^2/Q^2)$ from the massive
$B^{\nfive}_i(m_c^2/Q^2, m_b^2/Q^2)$, while
removing all the large logarithms $\log (m_b^2/Q^2)$, leaves untouched
$m_c$ dependent terms: this is fine since at large $Q^2$ such terms
are always suppressed by $m_c^2/Q^2$, so can never become large even
if enhanced by large logarithms $\log(m_b^2/Q^2)$.
Note that if there were no intrinsic charm, $C_j^{\nfour}(m_c^2/Q^2, m_b^2/Q^2)$
in Eq.~\eqref{eq:Bfourb} would be replaced by $\bar C_j^{\nfour}(m_c^2/Q^2, m_b^2/Q^2)$.
 
Of course the intrinsic beauty distribution must be very small indeed, suppressed by roughly $m_c^2/m_b^2$ compared to the intrinsic charm distribution. We can set it to zero by hand by taking as a boundary condition $f_b^{\nfour} = f_{\bar{b}}^{\nfour}=0$, or in the massless scheme with five active flavours at $\mu_b^2\sim m_b^2$ taking
\begin{equation}  
f_b^{\nfive} (\mu_b^2)  = \sum_{j,k = g,q', \bar{q}'} K^{\nfive}_{bj}\left( \smallfrac{m_b^2}{\mu_b^2}\right)\otimes
\tilde K{{}^{\nfive}_{jk}}^{-1} \left( \smallfrac{m_b^2}{\mu_b^2}\right) \otimes f_k^{\nfive} (\mu_b^2)
;\label{eq:zibbc}
\end{equation}
the last three lines in Eq.~(\ref{eq:FONLLB}) can then be dropped as they are subleading. Just as in Eq.~(\ref{eq:FONLLIC4}) we can write them in the compact form 
\begin{align}
\Delta F_{\rm FONLL} \left(Q^2,m_c^2,m_b^2  \right) &=  
 \sum_{i = b, \bar{b}}
  \Big[C_i^{\nfive}\left( \smallfrac{m_c^2}{Q^2},\smallfrac{m_b^2}{Q^2}\right)-C_i^{\nfive}\left( 0,0\right)\Big] \nonumber\\
&\qquad\qquad\otimes
\Big[f_i^{\nfive}  - \sum_{k,l = q', \bar{q'}, g} 
K_{ik}^{\nff}\otimes \tilde{K}{{}^{\nfive}_{kl}}^{-1}\otimes f_l^{\nfive}\Big].
\label{eq:FONLLIB5}
\end{align} 
From this we see that when there is no intrinsic beauty, in coefficient functions $C_{b,\bar{b}}^{\nfive}$ with an incoming bottom quark we can ignore both charm mass dependence and bottom mass dependence, treating both quarks as massless in these diagrams. We then get the S-ACOT expression for the structure function, corresponding to just the first two lines of Eq.~(\ref{eq:FONLLB}):
\beq\label{eq:S-ACOTb}
F_{\text{S-ACOT}}(Q^2,m_c^2,m_b^2) = \sum_{i=g,q',\bar q',b,\bar b}
\bar C^{\nfive}_i \(\smallfrac{m_c^2}{Q^2},\smallfrac{m_c^2}{Q^2}\) \otimes f^{\nfive}_i(Q^2)
\eeq
where
\beq\label{eq:BbarFONLL}
\bar C^{\nfive}_{i}\(\smallfrac{m_c^2}{Q^2},\smallfrac{m_b^2}{Q^2}\) = \begin{cases}
B^{\nfive}_i \left( \smallfrac{m_c^2}{Q^2},\smallfrac{m_b^2}{Q^2}\right)-B^{\nfivez}_i \left( 0,\smallfrac{m_b^2}{Q^2}\right)+C_i^{\nfive}(0,0),& \mbox{$i=g,q',\bar{q'}$},\cr
C_i^{\nfive}(0,0),& \mbox{$i=b,\bar{b}$}. \end{cases}
\eeq
This is in contrast to the charm case: when there is no intrinsic charm, we can ignore the charm mass dependence in the coefficient functions $C_{c,\bar{c}}^{\nfour}$ with incoming charm quark but not in principle the bottom mass dependence, arising through virtual loops.

\subsection{The top quark}
\label{sec:top}

The whole procedure described in Sect.~\ref{sec:beauty} can be repeated at the top threshold: here of
course it is clear that all the top quarks are generated
perturbatively, but the necessity to resum large logarithms of
$m_t^2/Q^2$ remains, and can be a performed by evolution of a top PDF
in a 6FS. Here S-ACOT corresponds to setting $m_c=m_b=m_t=0$ in all
diagrams with an incoming top quark: writing $q'' = q,c,b$ and
including explicit $m_t$ dependence in the coefficient functions, then
with the definitions
\bea
B^{\nsix}_i\left(\smallfrac{m_c^2}{Q^2}, \smallfrac{m_b^2}{Q^2}, \smallfrac{m_t^2}{Q^2}\right) &= 
\sum_{j = g, q'', \bar{q}''}
  \bar{C}_j^{\nfive} \left(\smallfrac{m_c^2}{Q^2}, \smallfrac{m_b^2}{Q^2},\smallfrac{m_t^2}{Q^2}\right)\otimes
  \tilde{K}{{}^{\nsf}_{ji}}^{-1} \(\smallfrac{m_t^2}{Q^2}\)
\label{eq:Bfivet}\\
B^{\nsixz}_i\left(0, 0,\smallfrac{m_t^2}{Q^2} \right) &= 
\sum_{j = g, q'', \bar{q}''}
  C_j^{\nfivez} \left(0,0, \smallfrac{m_t^2}{Q^2}\right)\otimes
  \tilde{K}{{}^{\nsf}_{ji}}^{-1} \(\smallfrac{m_t^2}{Q^2}\)
\label{eq:Bfivezt}\\
C_j^{\nfivez}\left(0,0, \smallfrac{m_t^2}{Q^2} \right)
&=\sum_{j = g, q'', \bar{q}'', t, \bar t} C_j^{\nsix}\left(0,0,0 \right)\otimes
  K_{ji}^{\nsf}\left( \smallfrac{m_t^2}{Q^2}\right),
\label{eq:C50}
\eea
we have 
\beq\label{eq:TbarFONLL}
\bar C^{\nsix}_{i}\(\smallfrac{m_c^2}{Q^2},\smallfrac{m_b^2}{Q^2},\smallfrac{m_t^2}{Q^2}\) = \begin{cases}
B^{\nsix}_i \left( \smallfrac{m_c^2}{Q^2},\smallfrac{m_b^2}{Q^2},\smallfrac{m_t^2}{Q^2}\right)-B^{\nsixz}_i \left( 0,0,\smallfrac{m_t^2}{Q^2}\right)+C_i^{\nsix}(0,0,0),& \mbox{$i=g,q'',\bar{q''}$},\cr
C_i^{\nsix}(0,0,0),& \mbox{$i=t,\bar{t}$}. \end{cases}
\eeq
These are the coefficient functions which enter the structure functions
\beq\label{eq:S-ACOTt}
F_{\text{S-ACOT}}(Q^2,m_c^2,m_b^2,m_t^2) = \sum_{i=g,q'',\bar q'',t,\bar t}
\bar C^{\nsix}_i \(\smallfrac{m_c^2}{Q^2},\smallfrac{m_c^2}{Q^2},\smallfrac{m_t^2}{Q^2}\) \otimes f^{\nsix}_i(Q^2)
\eeq
in the 6 flavour S-ACOT scheme.

\subsection{Variable Flavour Number Scheme}
\label{sec:vfns}

Putting all this together, we can now construct a variable flavour
number scheme with charm, bottom and top quarks. For definiteness we
will assume that both bottom and top are generated perturbatively,
while charm may have an intrinsic component, which is the setup that
is likely to be used when performing a PDF fit.
The generic structure function can then be written equivalently in
the various different $n_f$-flavour schemes, leading in principle to identical results (to all orders in $\as$).
In practice, at finite order, each of them is more appropriate for a specific range of scales.
In particular, the 3FS is appropriate only for $Q\sim m_c$,
the 4FS for $m_c\lesssim Q \lesssim m_b$
the 5FS for $m_b\lesssim Q \lesssim m_t$
and the 6FS for $Q \gtrsim m_t$. For scales above these ranges the results
of a finite order calculation will be spoiled by large unresummed logarithms.

Therefore, one can construct a variable flavour number scheme
using each result in its specific region of validity, using the heavy quark thresholds
to switch from one result to another:
\bea
F (Q^2,m_c^2, m_b^2, m_t^2) =
\begin{cases}
\displaystyle\sum_{i = g,q, \bar{q}, c ,\bar{c}}
C_i^{\nthree} \left( \smallfrac{m_c^2}{Q^2}, \smallfrac{m_b^2}{Q^2}, \smallfrac{m_t^2}{Q^2}\right)\otimes
  f_i^{\nthree} (Q^2) & \qquad\;\, Q^2 < \muc^2\\
\displaystyle\sum_{i = g,q, \bar{q}, c ,\bar{c}}
 C_i^{\nfour} \left( \smallfrac{m_c^2}{Q^2}, \smallfrac{m_b^2}{Q^2}, \smallfrac{m_t^2}{Q^2} \right)\otimes
  f_i^{\nfour} (Q^2) & \muc^2\leq Q^2 < \mub^2\\
\displaystyle\sum_{i = g,q, \bar{q}, c ,\bar{c}, b,\bar b}
\bar C_i^{\nfive} \left( \smallfrac{m_c^2}{Q^2}, \smallfrac{m_b^2}{Q^2}, \smallfrac{m_t^2}{Q^2} \right)\otimes
  f_i^{\nfive} (Q^2) & \mub^2\leq Q^2 < \mut^2\\
\displaystyle\sum_{i = g,q, \bar{q}, c ,\bar{c}, b,\bar b, t,\bar t}
\bar C_i^{\nsix} \left( \smallfrac{m_c^2}{Q^2}, \smallfrac{m_b^2}{Q^2}, \smallfrac{m_t^2}{Q^2} \right)\otimes
  f_i^{\nsix} (Q^2)  & \mut^2 \leq Q^2
\end{cases}
\label{eq:VFNS}
\eea
Notice that the sum in the 3FS runs also over the charm, to accomodate a possible intrinsic component,
while the bottom and top PDFs are generated perturbatively and therefore appear in 5FS and 6FS (bottom) and 6FS (top) only;
consistently, we have used the S-ACOT coefficient functions $\bar C^\nfive_i$ and $\bar C^\nsix_i$
in the 5FS and 6FS formulations.
In principle the thresholds $\mu_i$ do not need to coincide with
the analogous thresholds in the perturbative evolution of PDFs,
as pointed out e.g.\ in Refs.~\cite{Olness:2008px,Kusina:2013slm},
provided in both cases $\mu_i\sim m_i$, for $i=c,b,t$.
Note that for all practical purposes the result in the 3FS can be ignored,
since it is not needed to describe the structure function in the region of validity
of perturbative QCD, $Q\gtrsim 1$~GeV, in particular in the case of a fitted charm PDF.

It is useful to observe that in general discontinuities arise at the thresholds
when switching from one scheme to another: these are formally higher order effects,
and are ultimately due to interference between the coefficient functions and the matching functions $K_{ij}$
in the perturbative evolution of the PDFs. If a strict expansion in $\as$ in all the
contributions entering each formulation of Eq.~\eqref{eq:VFNS} is performed,
these higher order interference is eliminated and the VFNS is continuous at threshold~\cite{Bonvini:2015pxa}.

Keeping track of the mass dependence from all three heavy flavors
in each coefficient function of Eq.~\eqref{eq:VFNS}
is rather complicated in general, as one has to deal with calculations with three different masses.
The simplest implementation of the VFNS, which can be seen as the minimally
improved version of the zero-mass VFNS,
is to deal with only one mass scale at a time:
for instance in the 5FS the charm is considered massless and the top infinitely heavy.
The advantage of this simpler VFNS, which is
often used in practical applications,\footnote
{With some observable-dependent exceptions: if one considers for instance
bottom production in DIS, this could be described in a 4FS, but clearly keeping the exact bottom mass dependence.}
is its simplicity: the problem
with it is that there are uncontrolled approximations, with some of
the terms that are dropped (for example the effect of bottom quark
loops, appearing at $\Ord(\as^2)$, just below the bottom threshold, or
charm mass effects at around the bottom threshold appearing already at
LO with intrinsic charm, or $\Ord(\as)$ with perturbative charm) being
potentially quite significant.

At low orders in $\as$, including the full mass dependence is
straightforward.  At NNLO (i.e.\ $\Ord(\as^2)$) for contributions with
an incoming light parton, and NLO (i.e.\ $\Ord(\as)$) for diagrams
with an incoming heavy parton, the diagrams have at most one heavy
quark, so the contributions from charm, bottom and top can simply be
added. At the next order (N$^3$LO for contributions with an incoming
light parton, and NNLO for diagrams with an incoming heavy parton)
diagrams can contain two heavy quarks, possibly of different flavour,
and so here the various combinations must be added. The expressions
with two masses are already rather complicated (see e.g.~\cite{Ablinger:2012qj,Ablinger:2011pb,Blumlein:2014zxa}).
Diagrams with three different masses only occur at N$^4$LO (N$^3$LO for
an incoming heavy parton).
Of course the dependence one the top quark
mass at scales below or around the bottom mass, and on the charm mass at scales
of order the top mass, must both be so small that they can be ignored
for all practical purposes.

%% file: sections/sec-summary.tex
\section{Summary and Outlook}
\label{sec:conclusion}

We have described the construction of a VFNS for inclusive deep-inelastic processes to any order in perturbation theory, assuming as a starting point the existence of the $\overline{\rm MS}$ and decoupling (CWZ) renormalization schemes and the massless $\MSb$ factorization scheme.
We further assume the formal existence of all PDFs (whether corresponding to massless or massive partons) at all scales: thresholds are taken account of through the hard coefficient functions. The result we find is essentially unique to any given order in perturbation theory: in particular we obtain the same result from the ACOT procedure and from the FONLL procedure. The reason for this is clear: once the renormalization and factorization scheme is fixed, and thus the PDFs and their evolution, the coefficient functions, containing all the dependence on quark masses, must also be fixed uniquely order by order in perturbation theory. 

Of course assuming the formal existence of a charm (or indeed bottom or top) PDF at a given scale is not the same as assuming it is non-negligible, still less observable. The theoretical assumption that the PDFs vanish at threshold can always be implemented through imposition of a boundary condition on the perturbative evolution. This introduces a subleading ambiguity in the formalism, which (being subleading) is small but not resolvable at any given order in perturbation theory. It is only this ambiguity (and differences regarding the ordering of the perturbation expansion) that distinguishes phenomenologically the S-ACOT and FONLL schemes.

The way this ambiguity arises in the usual construction of a VFNS is through the addition of a charm quark PDF in the transition from the massive scheme (valid near threshold) to the massless scheme (valid far above it). This increases the size of the space of active partons: the matching matrix is then not square, and thus has no unique inverse. We have side stepped this ambiguity by adopting a slightly different procedure: we add the charm quark PDF by extending the space of light partons in the massive scheme (where the charm quark PDF does not mix with the light quark and gluon PDFs, and is thus decoupled), and only then match to the massless scheme. The matching matrix is then square, with a unique inverse, and the charm PDF below threshold can be interpreted as `intrinsic' charm. The limit of no instrinsic charm can then be taken a posteriori, as a theoretical assumption, or instead the intrinsic charm can be determined empirically through a PDF fit - `fitted charm'. The construction can be trivially extended to beauty and top.

Whether in practice one adopts an empirical procedure (determining the heavy quark distribution through a fit to data), or a theoretical prejudice (setting the heavy quark distribution in the massive scheme to zero) then depends critically on the heavy quark mass, and the precision of existing data.
For charm, the charm mass is sufficiently low (at around $1.3$ GeV), and the data are sufficiently precise (at the level of a few per cent), that the empirical approach may be necessary. For beauty it is probably at present still best to set the intrinsic distribution to zero, since measurements of a few per mille are out of reach. It is difficult to foresee the need for an intrinsic top distribution, since the top quark decays before there is time for nonperturbative effects to be significant. We hope to perform an empirical determination of intrinsic charm by fitting a charm PDF in the NNPDF formalism in the near future, as set out in Ref.~\cite{Ball:2015tna}.

\acknowledgments
We thank Valerio Bertone, Stefano Forte and Juan Rojo for discussions.
L.~R. and M.~B. are supported by an European Research Council Starting Grant `PDF4BSM'.
R.~D.~B. is funded by an STFC Consolidated Grant ST/J000329/1.

%% file: sections/sec-appendixmat.tex
\section{Inversion of Matching Matrices}
\label{sec-appendixmat}

Here we give various expressions useful for the inversion of block diagonal matrices, with particular application to the inversion of the matching matrix $K$ when restricted to light and heavy subspaces.

Consider a block diagonal matrix
\begin{equation}
\left( \begin{array}{cc}
A & B  \\
C & D  \end{array} \right)
\end{equation}
where $A$ is $m \times m$, $D$ is $n\times n$, and both are invertible with inverses $A^{-1}$ and $D^{-1}$, while $B$ is $m \times n$ and $C$ is $n \times m$. Then the inverse of this matrix is
\begin{equation}
\left( \begin{array}{cc}
A & B  \\
C & D  \end{array} \right)^{-1} = 
\left( \begin{array}{cc}
{E}^{-1} & -A^{-1}B{F}^{-1} \\
-D^{-1}C{E}^{-1} & {F}^{-1}  \end{array} \right),
\end{equation}
where
\begin{equation}
{E} = A - BD^{-1}C,\qquad {F} = D - CA^{-1}B.
\end{equation}
The following results are also useful:
\begin{equation}
A^{-1}B{F}^{-1}={E}^{-1}BD^{-1}, \qquad
D^{-1}C{E}^{-1}={F}^{-1}CA^{-1},
\end{equation}
and 
\begin{equation}
{E}^{-1} = A^{-1}+ A^{-1}B{F}^{-1}C A^{-1},\qquad
{F}^{-1} = D^{-1}+ D^{-1}C{E}^{-1}B D^{-1}.
\end{equation}

Applying these general results to the matching matrix $K_{ij}$, $i,j=g, q, \bar{q}, c, \bar{c}$, separated into light and heavy subspaces, thus with $m=7$ and $n=2$, we find that on the diagonal 
\begin{align}
 K^{-1}_{ij} 
 &=\tilde{K}^{-1}_{ij} +
 \sum_{k,l=g,q, \bar{q}}\sum_{m,n=c, \bar{c}}\tilde{K}^{-1}_{ik}\otimes K_{km}\otimes K^{-1}_{mn}\otimes K_{nl}\otimes\tilde{K}^{-1}_{lj} ,
  &&i,j=g, q, \bar{q}
 \label{eq:Kinvqq}\\
 K^{-1}_{ij} 
 &=\tilde{K}^{-1}_{ij}+
 \sum_{k,l=c, \bar{c}}\sum_{m,n=g,q, \bar{q}}\tilde{K}^{-1}_{ik}\otimes K_{km}\otimes K^{-1}_{mn}\otimes K_{nl}\otimes\tilde{K}^{-1}_{lj} ,
  &&i,j=c, \bar{c}
 \label{eq:Kinvcc}
 \end{align}
 while the off-diagonal mixing is given by 
\begin{align} 
K^{-1}_{ij} &= - \sum_{k=g,q, \bar{q}}\sum_{l=c, \bar{c}}\tilde{K}^{-1}_{ik}\otimes K_{kl}\otimes K^{-1}_{lj}\nonumber\\
 &=- \sum_{k=g,q, \bar{q}}\sum_{l=c, \bar{c}}K^{-1}_{ik}\otimes K_{kl}\otimes \tilde{K}^{-1}_{lj} ,&&i=g,q, \bar{q} , &&j=c, \bar{c}
\label{eq:Kinvqc}\\
K^{-1}_{ij} &= - \sum_{k=c, \bar{c}}\sum_{l=g,q, \bar{q}}\tilde{K}^{-1}_{ik}\otimes K_{kl}\otimes K^{-1}_{lj}\nonumber\\
 &=- \sum_{k=c, \bar{c}}\sum_{l=g,q, \bar{q}}K^{-1}_{ik}\otimes K_{kl}\otimes \tilde{K}^{-1}_{lj} , &&i=c, \bar{c} , &&j=g,q, \bar{q}\, .
 \label{eq:Kinvcq}
 \end{align}
In all these expressions the inverses of the matrices $\tilde{K}_{ij}$ are taken in the light and heavy subspaces, while those of ${K}_{ij}$ are taken in the full space: 
\begin{align}
\sum_{k= g, q, \bar{q}, c, \bar{c}} K^{-1}_{ik}K_{kj} 
 &=\sum_{k= g, q, \bar{q}, c, \bar{c}} K_{ik}K^{-1}_{kj} = \delta_{ij} , &&i,j=g,q,\bar{q},c, \bar{c}
 \label{eq:Kinv}\\
 \sum_{k= g, q, \bar{q}} \tilde{K}^{-1}_{ik}K_{kj} 
 &=\sum_{k= g, q, \bar{q}} K_{ik}\tilde{K}^{-1}_{kj} = \delta_{ij}, &&i,j=g,q,\bar{q}
 \label{eq:Ktilinvgg}\\
 \sum_{k= c, \bar{c}} \tilde{K}^{-1}_{ik}K_{kj} 
 &=\sum_{k= c, \bar{c}} K_{ik}\tilde{K}^{-1}_{kj} = \delta_{ij} , &&i,j=c, \bar{c}.
 \label{eq:Ktilinvcc}
\end{align}
Note in particular that if we restrict to a subspace, $K^{-1}$ is not the inverse of $K$, rather
\begin{align}
 \sum_{k= g, q, \bar{q}} K^{-1}_{ik}\big[K_{kj} - \sum_{m,n=c,\bar{c}}K_{km}\tilde{K}^{-1}_{mn}K_{mj}\big] 
&= \delta_{ij}, &&i,j=g,q,\bar{q}
 \label{eq:Kbarinvgg}\\
 \sum_{k= c, \bar{c}} \tilde{K}^{-1}_{ik}\big[K_{kj} - \sum_{m,n=g,q,\bar{q}}K_{km}{K}^{-1}_{mn}K_{mj}\big] 
&= \delta_{ij} , &&i,j=c, \bar{c}.
 \label{eq:Kbarinvcc}
\end{align} 

%% file: sections/sec-results.tex
\section{Explicit Results}
\label{sec:pheno}

We present in this Appendix explicit expressions for neutral current DIS in terms of primary
ingredients such as massive 3FS and massless 4FS coefficient functions.
We restrict our study to the charm structure functions
(defined for present purposes to be that part of the structure function in which the 
struck quark is a charm quark)
$F_2^c(Q^2,m_c^2)$, $F_L^c(Q^2,m_c^2)$ and $F_3^c(Q^2,m_c^2)$ since this is where
the effect of any intrinsic charm will be most visible: extension to
other structure functions (both neutral and charged current) is
straightforward.

\subsection{$F_2^c$ to NNLO}
\label{sec:f2c}

$F_2^c$ receives contributions from incoming charm quarks (starting at LO), incoming gluons (starting at NLO),
and incoming light quarks (starting at NNLO):
\begin{equation}
F_2^c(Q^2,m_c^2) = F_{2,c}^c(Q^2,m_c^2)+ F_{2,g}^c(Q^2,m_c^2) + \sum_{q} F_{2,q}^c(Q^2,m_c^2).
\label{eq:f2cdecomp}
\end{equation}
As explained in Sect.~\ref{sec:fonll}, each of these contributions can
be further decomposed into the contribution from standard FONLL
without intrinsic charm, computed using Eq.~(\ref{eq:FONLLzIC}), and
what may be thought of as an intrinsic charm correction, computed
using Eq.~(\ref{eq:FONLLIC}):
\begin{equation}
F_{2,i}^c(Q^2,m_c^2) = F_{2,i}^c(Q^2,m_c^2)\big|_{\rm zic} + \Delta F_{2,i}^c(Q^2,m_c^2)
\label{eq:f2cdecompdelta}
\end{equation}
with $i=c, g, q$.  The `zero intrinsic charm' contributions are then
given by (ignoring the damping factor discussed in Sect.~\ref{sec:damping})
\begin{align}
F_{2,c}^c(Q^2,m_c^2)\big|_{\rm zic}
&= C_{2,c}^{\nfour} \big( 0,\alpha_s \big)
\otimes f^{\nfour}_{c+}(Q^2) ,\label{eq:zICf2cc}\\
F_{2,g}^c(Q^2,m_c^2)\big|_{\rm zic}
&= \[B^{\nfour}_{2,g}\left(\smallfrac{m_c^2}{Q^2},\alpha_s\right)
  -B^{\nfourz}_{2,g}\left(\smallfrac{m_c^2}{Q^2},\alpha_s\right)
  + C_{2,g}^{\nfour} \big( 0,\alpha_s \big)
\]\otimes f^{\nfour}_g(Q^2) ,\label{eq:zICf2cg}\\
F_{2,q}^c(Q^2,m_c^2)\big|_{\rm zic}
&= \[B^{\nfour}_{2,q}\left(\smallfrac{m_c^2}{Q^2},\alpha_s\right)
  -B^{\nfourz}_{2,q}\left(\smallfrac{m_c^2}{Q^2},\alpha_s\right)
  +C_{2,q}^{\nfour} \big(0, \alpha_s \big)
\]\otimes f^{\nfour}_{q+}(Q^2) ,
\label{eq:zICf2cq}
\end{align} 
where
\begin{equation}
f^{\nfour}_{q+}(Q^2)=f^{\nfour}_q(Q^2)+f^{\nfour}_{\bar{q}}(Q^2),\qquad 
f^{\nfour}_{c+}(Q^2)=f^{\nfour}_c(Q^2)+f^{\nfour}_{\bar{c}}(Q^2) .
\label{eq:qcplus}
\end{equation}
In what follows we will expand each of these contributions out to NNLO
in $\as\equiv\alpha_s^{\nfour}(Q^2)$, using Eq.~(\ref{eq:matchalpha}), and
we will employ the following notation for the expansion of the various
coefficient and matching functions:
\begin{align}
C_{i}^{\nfour} \left( 0,\alpha_s \right) &=\sum^{\infty}_{p = 0}\big( \alpha_s^{\nfour} (Q^2) \big)^p C^{\nfour,p}_{i}(0) ,
\label{eq:C4exp}\\
C_{i}^{\nthree} \left( \smallfrac{m_c^2}{Q^2},\alpha_s \right) &=\sum^{\infty}_{p = 0}\big( \alpha_s^{\nfour} (Q^2) \big)^p C^{\nthree,p}_{i}\left(\smallfrac{m_c^2}{Q^2}\right) ,
\label{eq:Cexp}\\
C_{i}^{\nthreez} \left( \smallfrac{m_c^2}{Q^2},\alpha_s \right) &=\sum^{\infty}_{p = 0}\big( \alpha_s^{\nfour} (Q^2) \big)^p C^{\nthreez,p}_{i}\left(\smallfrac{m_c^2}{Q^2}\right) ,\label{eq:C0exp}\\
K_{ij}\left(\smallfrac{m_c^2}{Q^2}\right) &= \delta_{ij}+\sum^{\infty}_{p = 1}\big( \alpha_s^{\nfour} (Q^2) \big)^p K^p_{ij}\left(\smallfrac{m_c^2}{Q^2}\right) ,\label{eq:Kexp}
\end{align}
The expansions of ${K}_{ij}^{-1}$ and $\tilde {K}_{ij}^{-1}$ can be
straightforwardly related to the expansion of $K_{ij}$. Many of the
$\Ord(\as)$ coefficients vanish,
\begin{equation}
K^1_{qq}=K^1_{qg}=K^1_{gq}=K^1_{cq}=K^1_{qc}=0,
\end{equation}
(and all combinations with a quark replaced with an anti quark)
further simplifying the expansions. For FONLL-A we need only the
$\Ord(\as)$ contributions, while for FONLL-C (and B) we also need
the $\Ord(\as^2)$ contributions for a full implementation. All of
the coefficient functions and matching coefficients are known up to
$\Ord(\as^2)$, except for those whose second index is a heavy quark
which are known only up to $\Ord(\as)$.
In the following, we will omit the argument $m_c^2/Q^2$ in the expansion coefficients of $K_{ij}$,
while we will keep the argument in the coefficient functions for clarity;
we will also use $\as\equiv\alpha_s^{\nfour}(Q^2)$.

Expanding Eqs.~(\ref{eq:zICf2cc}--\ref{eq:zICf2cq}) to NNLO, we have 
\begin{align}
F_{2,c}^c(Q^2,m_c^2)\big|_{\rm zic}
&= \left[C_{2,c}^{\nfour,0}(0) + \as C_{2,c}^{\nfour,1}(0) + \as^2 C_{2,c}^{\nfour,2}(0)\right]
\otimes f^{\nfour}_{c+}(Q^2) ,\label{eq:zICf2ccexp}\\
F_{2,g}^c(Q^2,m_c^2) \big|_{\rm zic}
&= \as \Big[C_{2,g}^{\nthree,1}\left(\smallfrac{m_c^2}{Q^2}\right)
  -C_{2,g}^{\nthreez,1}\left(\smallfrac{m_c^2}{Q^2}\right)
+ C_{2,g}^{\nfour,1}(0) \Big]\otimes f^{\nfour}_g(Q^2)\nonumber\\
&+\as^2 \Big[C_{2,g}^{\nthree,2}\left(\smallfrac{m_c^2}{Q^2}\right)
 -C_{2,g}^{\nthreez,2}\left(\smallfrac{m_c^2}{Q^2}\right)
 + C_{2,g}^{\nfour,2}(0) \nonumber\\
&\qquad\qquad -\Big(C_{2,g}^{\nthree,1}\left(\smallfrac{m_c^2}{Q^2}\right)
  -C_{2,g}^{\nthreez,1}\left(\smallfrac{m_c^2}{Q^2}\right)\Big)\otimes {K}_{gg}^1
\Big]\otimes f^{\nfour}_g(Q^2) ,\label{eq:zICf2cgexp}\\
F_{2,q}^c(Q^2,m_c^2) \big|_{\rm zic}
&= \as^2 \Big[C_{2,q}^{\nthree,2}\left(\smallfrac{m_c^2}{Q^2}\right)
 -C_{2,q}^{\nthreez,2}\left(\smallfrac{m_c^2}{Q^2}\right)
 + C_{2,q}^{\nfour,2}(0) 
\Big]\otimes f^{\nfour}_{q+}(Q^2) ,\label{eq:zICf2cqexp}
\end{align}
where we also expanded $B^{\nfour}_{2,g}$ (and its massless limit) using the definition Eq.~(\ref{eq:Bdef}).

We now focus on the new terms $\Delta F_{2,i}^c(Q^2,m_c^2)$: using
Eq.~\eqref{eq:FONLLIC}, and dropping terms which do not contribute at
NNLO, we have
\begin{align}
\Delta F_{2,c}^c(Q^2,m_c^2)
&= \sum_{i=c,\bar c}\Big[C_{2,i}^{\nthree}\left(\smallfrac{m_c^2}{Q^2},\as\right) 
- C_{2,i}^{\nthreez}\left(\smallfrac{m_c^2}{Q^2},\as\right) \Big]\otimes K^{-1}_{ic}\otimes f^{\nfour}_{c+}(Q^2) \nonumber\\
&\quad- \Big[ C_{2,g}^{\nthree}\left(\smallfrac{m_c^2}{Q^2},\as\right) 
- C_{2,g}^{\nthreez}\left(\smallfrac{m_c^2}{Q^2},\as\right)\Big]
\otimes K_{gc}\otimes f^{\nfour}_{c+}(Q^2)
+\Ord(\alpha_s^3) ,\label{eq:ICf2cc}\\
\Delta F_{2,g}^c(Q^2,m_c^2)
&= -2\Big[C_{2,c}^{\nthree}\left(\smallfrac{m_c^2}{Q^2},\as\right) 
- C_{2,c}^{\nthreez}\left(\smallfrac{m_c^2}{Q^2},\as\right)\Big]\nonumber\\ &\qquad\qquad\qquad\qquad
\otimes {K}^{-1}_{cc}\otimes {K}_{cg}\otimes {K}^{-1}_{gg}\otimes f^{\nfour}_{g}(Q^2)+\Ord(\alpha_s^3) ,\label{eq:ICf2cg}\\
\Delta F_{2,q}^c(Q^2,m_c^2)
&= -2\Big[C_{2,c}^{\nthree}\left(\smallfrac{m_c^2}{Q^2},\as\right) - C_{2,c}^{\nthreez}\left(\smallfrac{m_c^2}{Q^2},\as\right)\Big] 
\otimes  K_{cq}\otimes f^{\nfour}_{q+}(Q^2)+\Ord(\alpha_s^3) ,\label{eq:ICf2cq}
\end{align} 
where the second line in Eq.~\eqref{eq:ICf2cc} comes from the $B$ terms in Eq.~\eqref{eq:FONLLIC}
which do not contribute at this order in the other two cases.
Expanding to NNLO we find
\begin{align}
\Delta F_{2,c}^c(Q^2,m_c^2)
&= \Big[C_{2,c}^{\nthree,0}\left(\smallfrac{m_c^2}{Q^2}\right) 
- C_{2,c}^{\nthreez,0}\left(\smallfrac{m_c^2}{Q^2}\right)\Big]\otimes f^{\nfour}_{c+}(Q^2) \nonumber\\
&+ \as\Big[C_{2,c}^{\nthree,1}\left(\smallfrac{m_c^2}{Q^2}\right) 
- C_{2,c}^{\nthreez,1}\left(\smallfrac{m_c^2}{Q^2}\right)\nonumber\\
&\qquad - \(C_{2,c}^{\nthree,0}\left(\smallfrac{m_c^2}{Q^2}\right) 
- C_{2,c}^{\nthreez,0}\left(\smallfrac{m_c^2}{Q^2}\right)\)\otimes K_{cc}^1\Big]\otimes f^{\nfour}_{c+}(Q^2)\nonumber\\
&+ \as^2\Big[C_{2,c}^{\nthree,2}\left(\smallfrac{m_c^2}{Q^2}\right) 
- C_{2,c}^{\nthreez,2}\left(\smallfrac{m_c^2}{Q^2}\right)\nonumber\\
&\qquad - \(C_{2,c}^{\nthree,1}\left(\smallfrac{m_c^2}{Q^2}\right) 
- C_{2,c}^{\nthreez,1}\left(\smallfrac{m_c^2}{Q^2}\right)\)\otimes K_{cc}^1
\nonumber\\
&\qquad - \(C_{2,c}^{\nthree,0}\left(\smallfrac{m_c^2}{Q^2}\right) 
- C_{2,c}^{\nthreez,0}\left(\smallfrac{m_c^2}{Q^2}\right)\)\otimes \(K_{cc}^2+K_{\bar cc}^2 - 2{K}_{cg}^1 \otimes {K}_{gc}^1-{K}_{cc}^1 \otimes {K}_{cc}^1 \)
\nonumber\\
&\qquad - \(C_{2,g}^{\nthree,1}\left(\smallfrac{m_c^2}{Q^2}\right) 
- C_{2,g}^{\nthreez,1}\left(\smallfrac{m_c^2}{Q^2}\right)\)\otimes K_{gc}^1
\Big]\otimes f^{\nfour}_{c+}(Q^2) ,
\label{eq:Deltaf2cc}\\
\Delta F_{2,g}^c(Q^2,m_c^2)
&=  - \as2\Big[C_{2,c}^{\nthree,0}\left(\smallfrac{m_c^2}{Q^2}\right) 
- C_{2,c}^{\nthreez,0}\left(\smallfrac{m_c^2}{Q^2}\right)\Big]\otimes K_{cg}^1\otimes f^{\nfour}_{g}(Q^2)\nonumber\\
&- \as^2 2\Big[\(C_{2,c}^{\nthree,1}\left(\smallfrac{m_c^2}{Q^2}\right) 
- C_{2,c}^{\nthreez,1}\left(\smallfrac{m_c^2}{Q^2}\right)\)\otimes K_{cg}^1\nonumber\\
&\qquad +\(C_{2,c}^{\nthree,0}\left(\smallfrac{m_c^2}{Q^2}\right) 
- C_{2,c}^{\nthreez,0}\left(\smallfrac{m_c^2}{Q^2}\right)\)\otimes \(K_{cg}^2 -K_{cg}^1\otimes K_{gg}^1-K_{cc}^1\otimes K_{cg}^1 \)
\Big]\nonumber\\
&\qquad\otimes f^{\nfour}_{g}(Q^2) ,\label{eq:Deltaf2cg}\\
\Delta F_{2,q}^c(Q^2,m_c^2)
&=  - \as^2 2\Big[C_{2,c}^{\nthree,0}\left(\smallfrac{m_c^2}{Q^2}\right) 
- C_{2,c}^{\nthreez,0}\left(\smallfrac{m_c^2}{Q^2}\right)\Big]\otimes K_{cq}^2\otimes f^{\nfour}_{q+}(Q^2) .\label{eq:Deltaf2cq}
\end{align} 
Note that the LO coefficient functions are proportional to
$\delta$-functions, so their convolutions with the matching terms are
trivial. While the LO and NLO terms can all be computed
in full, many of the other terms at NNLO cannot yet be computed since
the NNLO diagrams with an incoming heavy quark line have yet to be
evaluated.

Using Eq.~\eqref{eq:masslimdef}, we can re-express the coefficients $C_{2,i}^{\nthreez,k}$
appearing in the previous equations in terms
of the expansion coefficients of $C^{\nfour}_{2,i}$ and $K_{ij}$.
Up to $\Ord(\as^2)$ we have
\bea
C^{\nthreez}_{2,c}\(\smallfrac{m_c^2}{Q^2}\) &= C^{\nfour,0}_{2,c}(0)
+\as\[ C^{\nfour,1}_{2,c}(0) +C^{\nfour,0}_{2,c}(0)\otimes K_{cc}^1 \]\nonumber\\
&+\as^2\[ C^{\nfour,2}_{2,c}(0) +C^{\nfour,1}_{2,c}(0)\otimes K_{cc}^1 +C^{\nfour,1}_{2,g}(0)\otimes K_{gc}^1
+C^{\nfour,0}_{2,c}(0)\otimes (K_{cc}^2+ K_{\bar cc}^2)
 \]\nonumber\\
&+\Ord(\as^3),
\label{eq:C30c}
\\
C^{\nthreez}_{2,g}\(\smallfrac{m_c^2}{Q^2}\)
&= \as\[ C^{\nfour,1}_{2,g}(0) + 2C^{\nfour,0}_{2,c}(0)\otimes K_{cg}^1 \]\nonumber\\
&+\as^2\[ C^{\nfour,2}_{2,g}(0) +C^{\nfour,1}_{2,g}(0)\otimes K_{gg}^1 +2C^{\nfour,1}_{2,c}(0)\otimes K_{cg}^1
+2C^{\nfour,0}_{2,c}(0)\otimes K_{cg}^2
 \]\nonumber\\
&+\Ord(\as^3),
\label{eq:C30g}
\\
C^{\nthreez}_{2,q}\(\smallfrac{m_c^2}{Q^2}\)
&=\as^2\[ C^{\nfour,2}_{2,q}(0) +2C^{\nfour,0}_{2,c}(0)\otimes K_{cq}^2\] +\Ord(\as^3).
\label{eq:C30q}
\eea
Substituting these into Eqs.~(\ref{eq:zICf2ccexp}--\ref{eq:Deltaf2cq}) we obtain expressions
for $F^c_{2,i}\big|_{\rm zic}$ and $\Delta F^c_{2,i}$ in terms of the expansion coefficients of
$C^{\nthree}_i(m_c^2/Q^2)$, $C^{\nfour}_i(0)$ and $K_{ij}(m_c^2/Q^2)$, which can be regarded as primary quantities.

When we evaluate the sum
$F_{2,c}^c(Q^2,m_c^2)\big|_{\rm zic} + \Delta F_{2,c}^c(Q^2,m_c^2)$ there are
considerable cancellations between the two terms: in fact we find
using Eq.~(\ref{eq:FONLLsimp})
\begin{align}
F_{2,c}^c(Q^2,m_c^2)
&= C_{2,c}^{\nthree,0}\left(\smallfrac{m_c^2}{Q^2}\right) 
\otimes f^{\nfour}_{c+}(Q^2) \nonumber\\
&+ \as\Big[C_{2,c}^{\nthree,1}\left(\smallfrac{m_c^2}{Q^2}\right) 
 - C_{2,c}^{\nthree,0}\left(\smallfrac{m_c^2}{Q^2}\right) 
\otimes K_{cc}^1\Big]\otimes f^{\nfour}_{c+}(Q^2)\nonumber\\
&+ \as^2\Big[C_{2,c}^{\nthree,2}\left(\smallfrac{m_c^2}{Q^2}\right) 
 - C_{2,c}^{\nthree,1}\left(\smallfrac{m_c^2}{Q^2}\right) \otimes K_{cc}^1
\nonumber\\
&\qquad - C_{2,c}^{\nthree,0}\left(\smallfrac{m_c^2}{Q^2}\right)\otimes \(K_{cc}^2+ K_{\bar cc}^2-2K_{cg}^1 \otimes K_{gc}^1-K_{cc}^1 \otimes K_{cc}^1\)
\nonumber\\
&\qquad -  C_{2,g}^{\nthree,1}\left(\smallfrac{m_c^2}{Q^2}\right) \otimes K_{gc}^1
\Big]\otimes f^{\nfour}_{c+}(Q^2) ,\label{eq:f2ccsimp}\\
F_{2,g}^c(Q^2,m_c^2)
&=  \as \Big[C_{2,g}^{\nthree,1}\left(\smallfrac{m_c^2}{Q^2}\right)
- 2C_{2,c}^{\nthree,0}\left(\smallfrac{m_c^2}{Q^2}\right) 
\otimes K_{cg}^1\Big]\otimes f^{\nfour}_{g}(Q^2)\nonumber\\
&+\as^2 \Big[
C_{2,g}^{\nthree,2}\left(\smallfrac{m_c^2}{Q^2}\right)
- C_{2,g}^{\nthree,1}\left(\smallfrac{m_c^2}{Q^2}\right) 
\otimes K_{gg}^1
- 2C_{2,c}^{\nthree,1}\left(\smallfrac{m_c^2}{Q^2}\right) \otimes K_{cg}^1\nonumber\\
&\qquad-2C_{2,c}^{\nthree,0}\left(\smallfrac{m_c^2}{Q^2}\right)\otimes \(K_{cg}^2
-K_{cg}^1\otimes K_{gg}^1-K_{cc}^1\otimes K_{cg}^1\)\Big]\otimes f^{\nfour}_{g}(Q^2) ,\label{eq:f2cgsimp}\\
F_{2,q}^c(Q^2,m_c^2)
&=  \as^2 \Big[C_{2,q}^{\nthree,2}\left(\smallfrac{m_c^2}{Q^2}\right) -2C_{2,c}^{\nthree,0}\left(\smallfrac{m_c^2}{Q^2}\right) 
\otimes K_{cq}^2\Big] \otimes f^{\nfour}_{q+}(Q^2) .\label{eq:f2cqsimp}
\end{align}
Note that we can choose to order these expansions in different ways:
the FONLL scheme naming FONLL-A includes all terms to
$\Ord(\alpha_s)$, FONLL-C includes all terms to $\Ord(\alpha_s^2)$,
while in FONLL-B one adds the logarithmic parts of the $O(\alpha_s^2)$
contributions not originated by a charm quark to FONLL-A. Note however
that the massive NNLO coefficients $C_{2,c}^{\nthree,2}(m_c^2/Q^2)$
and the NNLO matching functions $K_{ic}^2$ are not known: for this
reason FONLL-B and FONLL-C cannot be fully computed at present if
one has to account for a possible intrinsic charm.  The best option
for going beyond NLO is to use full FONLL-A plus the `zero intrinsic
charm' contribution at higher orders, the $\Delta F$ terms
being set to zero beyond $\Ord(\as)$.  Alternatively, as proposed
in Ref.~\cite{Stavreva:2012bs}, it is possible to use a $\chi$
rescaling on the massless NNLO charm initiated coefficients to mimic
the dominant charm mass effects.

\subsection{$F_L^c$ to NNLO}
\label{sec:fLc}

Precisely the same arguments can be used for other structure functions, in particular $F_L^c$ and $F_3^c$. Writing
\begin{equation}
F_L^c(Q^2,m_c^2) = F_{L,c}^c(Q^2,m_c^2)+ F_{L,g}^c(Q^2,m_c^2) + \sum_q F_{L,q}^c(Q^2,m_c^2).
\label{eq:fLcdecomp}
\end{equation}
we find for the zero intrinsic charm contributions
\begin{align}
F_{L,c}^c(Q^2,m_c^2)\big|_{\rm zic}
&= \Big[\as C_{L,c}^{\nfour,1}(0) + \as^2 C_{L,c}^{\nfour,2}(0)\Big]
\otimes f^{\nfour}_{c+}(Q^2) ,\label{eq:zICfLcc}\\
F_{L,g}^c(Q^2,m_c^2) \big|_{\rm zic}
&= \as \Big[C_{L,g}^{\nthree,1}\left(\smallfrac{m_c^2}{Q^2}\right)
  -C_{L,g}^{\nthreez,1}\left(\smallfrac{m_c^2}{Q^2}\right)
+ C_{L,g}^{\nfour,1}(0) \Big]\otimes f^{\nfour}_g(Q^2)\nonumber\\
&+\as^2 \Big[C_{L,g}^{\nthree,2}\left(\smallfrac{m_c^2}{Q^2}\right)
 -C_{L,g}^{\nthreez,2}\left(\smallfrac{m_c^2}{Q^2}\right)
 + C_{L,g}^{\nfour,2}(0) \nonumber\\
&\qquad\qquad -\Big(C_{L,g}^{\nthree,1}\left(\smallfrac{m_c^2}{Q^2}\right)
  -C_{L,g}^{\nthreez,1}\left(\smallfrac{m_c^2}{Q^2}\right)\Big)\otimes {K}_{gg}^1
\Big]\otimes f^{\nfour}_g(Q^2) ,\label{eq:zICfLcg}\\
F_{L,q}^c(Q^2,m_c^2) \big|_{\rm zic}
&= \as^2 \Big[C_{L,q}^{\nthree,2}\left(\smallfrac{m_c^2}{Q^2}\right)
 -C_{L,q}^{\nthreez,2}\left(\smallfrac{m_c^2}{Q^2}\right)
 + C_{L,q}^{\nfour,2} (0)
\Big]\otimes f^{\nfour}_{q+}(Q^2) ,\label{eq:zICfLcq}
\end{align}
while
\begin{align}
\Delta F_{L,c}^c(Q^2,m_c^2)
&= C_{L,c}^{\nthree,0}\left(\smallfrac{m_c^2}{Q^2}\right)\otimes f^{\nfour}_{c+}(Q^2) \nonumber\\
&+ \as\Big[C_{L,c}^{\nthree,1}\left(\smallfrac{m_c^2}{Q^2}\right) 
  - C_{L,c}^{\nthreez,1}\left(\smallfrac{m_c^2}{Q^2}\right)
  - C_{L,c}^{\nthree,0}\left(\smallfrac{m_c^2}{Q^2}\right)\otimes K_{cc}^1\Big]\otimes f^{\nfour}_{c+}(Q^2)\nonumber\\
&+ \as^2\Big[C_{L,c}^{\nthree,2}\left(\smallfrac{m_c^2}{Q^2}\right) 
- C_{L,c}^{\nthreez,2}\left(\smallfrac{m_c^2}{Q^2}\right)\nonumber\\
&\qquad - \(C_{L,c}^{\nthree,1}\left(\smallfrac{m_c^2}{Q^2}\right) 
- C_{L,c}^{\nthreez,1}\left(\smallfrac{m_c^2}{Q^2}\right)\)\otimes K_{cc}^1
\nonumber\\
&\qquad - C_{L,c}^{\nthree,0}\left(\smallfrac{m_c^2}{Q^2}\right)\otimes
\(K_{cc}^2+ K_{\bar cc}^2 - 2K_{cg}^1 \otimes {K}_{gc}^1 - {K}_{cc}^1 \otimes {K}_{cc}^1\)
\nonumber\\
&\qquad - \(C_{L,g}^{\nthree,1}\left(\smallfrac{m_c^2}{Q^2}\right) 
- C_{L,g}^{\nthreez,1}\left(\smallfrac{m_c^2}{Q^2}\right)\)\otimes K_{gc}^1
\Big]\otimes f^{\nfour}_{c+}(Q^2) ,\label{eq:DeltafLcc}\\
\Delta F_{L,g}^c(Q^2,m_c^2)
&=  - \as 2 C_{L,c}^{\nthree,0}\left(\smallfrac{m_c^2}{Q^2}\right) 
\otimes K_{cg}^1\otimes f^{\nfour}_{g}(Q^2)\nonumber\\
&- \as^2 2\Big[\(C_{L,c}^{\nthree,1}\left(\smallfrac{m_c^2}{Q^2}\right) 
- C_{L,c}^{\nthreez,1}\left(\smallfrac{m_c^2}{Q^2}\right)\)\otimes K_{cg}^1\nonumber\\
&\qquad +C_{L,c}^{\nthree,0}\left(\smallfrac{m_c^2}{Q^2}\right)\otimes
  \(K_{cg}^2-K_{cg}^1\otimes K_{gg}^1-K_{cc}^1\otimes K_{cg}^1\)\Big]\otimes f^{\nfour}_{g}(Q^2) ,\label{eq:DeltafLcg}\\
\Delta F_{L,q}^c(Q^2,m_c^2)
&=  - \as^2 2C_{L,c}^{\nthree,0}\left(\smallfrac{m_c^2}{Q^2}\right) 
\otimes K_{cq}^2\otimes f^{\nfour}_{q+}(Q^2) .\label{eq:DeltafLcq}
\end{align}
Note that although $C^{\nthreez,0}_{L,c}=0$, $C^{\nthree,0}_{L,c}(m_c^2/Q^2)$ is nontrivial,
though power suppressed.
One can use expressions similar to Eqs.~(\ref{eq:C30c}--\ref{eq:C30q}) to re-express the structure functions
above in terms of only $C^{\nthree}_{L,i}$, $C^{\nfour}_{L,i}$ and $K_{ij}$.
When we evaluate the sum $F_{L,c}^c(Q^2,m_c^2)\big|_{\rm zic} + \Delta F_{L,c}^c(Q^2,m_c^2)$  we find 
\begin{align}
F_{L,c}^c(Q^2,m_c^2)
&= C_{L,c}^{\nthree,0}\left(\smallfrac{m_c^2}{Q^2}\right) 
\otimes f^{\nfour}_{c+}(Q^2) \nonumber\\
&+ \as\Big[C_{L,c}^{\nthree,1}\left(\smallfrac{m_c^2}{Q^2}\right) 
 - C_{L,c}^{\nthree,0}\left(\smallfrac{m_c^2}{Q^2}\right) 
\otimes K_{cc}^1\Big]\otimes f^{\nfour}_{c+}(Q^2)\nonumber\\
&+ \as^2\Big[C_{L,c}^{\nthree,2}\left(\smallfrac{m_c^2}{Q^2}\right) 
 - C_{L,c}^{\nthree,1}\left(\smallfrac{m_c^2}{Q^2}\right) \otimes K_{cc}^1
\nonumber\\
&\qquad - C_{L,c}^{\nthree,0}\left(\smallfrac{m_c^2}{Q^2}\right)\otimes \(K_{cc}^2+ K_{\bar cc}^2-2K_{cg}^1 \otimes K_{gc}^1-K_{cc}^1 \otimes K_{cc}^1\)
\nonumber\\
&\qquad  - C_{L,g}^{\nthree,1}\left(\smallfrac{m_c^2}{Q^2}\right) \otimes K_{gc}^1
\Big]\otimes f^{\nfour}_{c+}(Q^2) ,\label{eq:fLccsimp}\\
F_{L,g}^c(Q^2,m_c^2)
&=  \as \Big[C_{L,g}^{\nthree,1}\left(\smallfrac{m_c^2}{Q^2}\right)
- 2C_{L,c}^{\nthree,0}\left(\smallfrac{m_c^2}{Q^2}\right) 
\otimes K_{cg}^1\Big]\otimes f^{\nfour}_{g}(Q^2)\nonumber\\
&+\as^2 \Big[
C_{L,g}^{\nthree,2}\left(\smallfrac{m_c^2}{Q^2}\right)
- C_{L,g}^{\nthree,1}\left(\smallfrac{m_c^2}{Q^2}\right) 
\otimes K_{gg}^1
- 2C_{L,c}^{\nthree,1}\left(\smallfrac{m_c^2}{Q^2}\right) \otimes K_{cg}^1\nonumber\\
&\qquad-2C_{L,c}^{\nthree,0}\left(\smallfrac{m_c^2}{Q^2}\right)\otimes 
\(K_{cg}^2 - K_{cg}^1\otimes K_{gg}^1 - K_{cc}^1\otimes K_{cg}^1\)\Big]\otimes f^{\nfour}_{g}(Q^2) ,\label{eq:fLcgsimp}\\
F_{L,q}^c(Q^2,m_c^2)
&=  \as^2 \Big[C_{L,q}^{\nthree,2}\left(\smallfrac{m_c^2}{Q^2}\right) -2C_{L,c}^{\nthree,0}\left(\smallfrac{m_c^2}{Q^2}\right) 
\otimes K_{cq}^2\Big] \otimes f^{\nfour}_{q+}(Q^2) .\label{eq:fLcqsimp}
\end{align}

\subsection{$F_3^c$ to NNLO}
\label{sec:f3c}

Similarly we write
\begin{equation}
F_3^c(Q^2,m_c^2) = F_{3,c}^c(Q^2,m_c^2)+ \sum_q F_{3,q}^c(Q^2,m_c^2)
\label{eq:f3cdecomp}
\end{equation}
and 
\begin{equation}
f^{\nfour}_{q-}(Q^2)=f^{\nfour}_q(Q^2)-f^{\nfour}_{\bar{q}}(Q^2),\qquad 
f^{\nfour}_{c-}(Q^2)=f^{\nfour}_c(Q^2)-f^{\nfour}_{\bar{c}}(Q^2) .
\label{eq:qcminus}
\end{equation}
At NLO the valence distributions $f^{\nfour}_{c-}$ and $f^{\nfour}_{q-}$ evolve independently, but at  NNLO they mix. Thus even if there is no intrinsic charm, a small valence charm distribution $f^{\nfour}_{c-}$ will be generated dynamically above threshold. We thus find that
\begin{align}
F_{3,c}^c(Q^2,m_c^2)\big|_{\rm zic}
&= \Big[C_{3,c}^{\nfour,0}(0) + \as C_{3,c}^{\nfour,1}(0) + \as^2 C_{3,c}^{\nfour,2}(0)\Big]
\otimes f^{\nfour}_{c-}(Q^2) ,\label{eq:zICf3cc}\\
F_{3,q}^c(Q^2,m_c^2)\big|_{\rm zic}
&= \as^2 \Big[C_{3,q}^{\nthree,2}\left(\smallfrac{m_c^2}{Q^2}\right)
 -C_{3,q}^{\nthreez,2}\left(\smallfrac{m_c^2}{Q^2}\right)
 + C_{3,q}^{\nfour,2} (0)
\Big]\otimes f^{\nfour}_{q-}(Q^2) ,\label{eq:zICf3cq}
\end{align}
while 
\begin{align}
\Delta F_{3,c}^c(Q^2,m_c^2)
&= \Big[C_{3,c}^{\nthree,0}\left(\smallfrac{m_c^2}{Q^2}\right) 
- C_{3,c}^{\nthreez,0}\left(\smallfrac{m_c^2}{Q^2}\right)\Big]\otimes f^{\nfour}_{c-}(Q^2) \nonumber\\
&+ \as\Big[C_{3,c}^{\nthree,1}\left(\smallfrac{m_c^2}{Q^2}\right) 
- C_{3,c}^{\nthreez,1}\left(\smallfrac{m_c^2}{Q^2}\right)\nonumber\\
&\qquad - \(C_{3,c}^{\nthree,0}\left(\smallfrac{m_c^2}{Q^2}\right) 
- C_{3,c}^{\nthreez,0}\left(\smallfrac{m_c^2}{Q^2}\right)\)\otimes K_{cc}^1\Big]\otimes f^{\nfour}_{c-}(Q^2)\nonumber\\
&+ \as^2\Big[C_{3,c}^{\nthree,2}\left(\smallfrac{m_c^2}{Q^2}\right) 
- C_{3,c}^{\nthreez,2}\left(\smallfrac{m_c^2}{Q^2}\right)\nonumber\\
&\qquad - \(C_{3,c}^{\nthree,1}\left(\smallfrac{m_c^2}{Q^2}\right) 
- C_{3,c}^{\nthreez,1}\left(\smallfrac{m_c^2}{Q^2}\right)\)\otimes K_{cc}^1
\nonumber\\
&\qquad - \(C_{3,c}^{\nthree,0}\left(\smallfrac{m_c^2}{Q^2}\right) 
- C_{3,c}^{\nthreez,0}\left(\smallfrac{m_c^2}{Q^2}\right)\)
\nonumber\\
&\qquad\qquad \otimes (K_{cc}^2+ K_{\bar cc}^2-2K_{cg}^1 \otimes K_{gc}^1-K_{cc}^1 \otimes K_{cc}^1)
\Big]\otimes f^{\nfour}_{c-}(Q^2) ,\label{eq:Deltaf3cc}\\
\Delta F_{3,q}^c(Q^2,m_c^2)
&=  - 2\as^2 \Big[C_{3,c}^{\nthree,0}\left(\smallfrac{m_c^2}{Q^2}\right) 
- C_{3,c}^{\nthreez,0}\left(\smallfrac{m_c^2}{Q^2}\right)\Big]\otimes K_{cq}^2\otimes f^{\nfour}_{q-}(Q^2) .\label{eq:Deltaf3cq}
\end{align}
Now when we evaluate the sum $F_{3,c}^c(Q^2,m_c^2\big|_{\rm zic} + \Delta F_{3,c}^c(Q^2,m_c^2)$ we have
\begin{align}
F_{3,c}^c(Q^2,m_c^2)
&= C_{3,c}^{\nthree,0}\left(\smallfrac{m_c^2}{Q^2}\right) 
\otimes f^{\nfour}_{c-}(Q^2) \nonumber\\
&+ \as\Big[C_{3,c}^{\nthree,1}\left(\smallfrac{m_c^2}{Q^2}\right) 
 - C_{3,c}^{\nthree,0}\left(\smallfrac{m_c^2}{Q^2}\right) 
\otimes K_{cc}^1\Big]\otimes f^{\nfour}_{c-}(Q^2)\nonumber\\
&+ \as^2\Big[C_{3,c}^{\nthree,2}\left(\smallfrac{m_c^2}{Q^2}\right) 
 - C_{3,c}^{\nthree,1}\left(\smallfrac{m_c^2}{Q^2}\right) \otimes K_{cc}^1
\nonumber\\
&\qquad - C_{3,c}^{\nthree,0}\left(\smallfrac{m_c^2}{Q^2}\right)\otimes (K_{cc}^2+ K_{\bar cc}^2-2K_{cg}^1 \otimes K_{gc}^1-K_{cc}^1 \otimes K_{cc}^1)
\Big]\otimes f^{\nfour}_{c-}(Q^2) ,\label{eq:f3ccsimp}\\
F_{3,q}^c(Q^2,m_c^2)
&=  \as^2 \Big[C_{3,q}^{\nthree,2}\left(\smallfrac{m_c^2}{Q^2}\right) -2C_{3,c}^{\nthree,0}\left(\smallfrac{m_c^2}{Q^2}\right) 
\otimes K_{cq}^2\Big] \otimes f^{\nfour}_{q-}(Q^2) .\label{eq:f3cgsimp}
\end{align}